# Cassini observations of Saturn's southern polar cusp

C.S. Arridge[1], J.M. Jasinski[2,3*], N. Achilleos[4,3], Y.V. Bogdanova[5], E.J. Bunce[6], S.W.H. Cowley[7], A.N. Fazakerley[2], K.K. Khurana[8,9], L. Lamy[10], J.S. Leisner[11], E. Roussos[12], C.T. Russell[8,9], P. Zarka[9], A.J. Coates[2,3], M.K. Dougherty[13], G.H. Jones[2,3], S.M. Krimigis[14], N. Krupp[12]

1. Department of Physics, Lancaster University, Bailrigg, Lancaster, LA1 4YB, United Kingdom.

2. Mullard Space Science Laboratory, University College London, Holmbury St. Mary, Dorking, Surrey, RH5 6NT, United Kingdom.

3. The Centre for Planetary Sciences at UCL/Birkbeck, Gower Street, London, WC1E 6BT, United Kingdom.

4. Department of Physics and Astronomy, University College London, Gower Street, London, WC1E 6BT, United Kingdom.

5. Rutherford Appleton Laboratory, Science and Technology Facilities Council, Harwell Oxford, Didcot, Oxon, OX11 0QX, United Kingdom.

6. Space Research Centre, Department of Physics and Astronomy, University of Leicester, Leicester, LE1 7RH, United Kingdom.

7. Department of Physics and Astronomy, University of Leicester, University Road, Leicester, LE1 7RH, United Kingdom.

8. Institute of Geophysics and Planetary Physics, University of California Los Angeles, Los Angeles, California, 90024-1567, United States of America.

9. Department of Earth and Space Sciences, University of California Los Angeles, Los Angeles, California, 90024-1567, United States of America.

10. LESIA, Observatoire de Paris, 92195, Meudon, France.

11. Arctic Slope Technical Services, Beltsville, Maryland, , United States of America.

12. Max-Planck-Institut für Sonnensystemforschung, Katlenburg-Lindau, Germany.

13. The Blackett Laboratory, Imperial College, South Kensington, London, SW7 2AZ, United Kingdom.

14. Johns Hopkins Applied Physics Laboratory, Laurel, MD, United States of America.

* Now at Climate and Space Sciences and Engineering, Space Research Building, University of Michigan, 2455 Hayward Street, Ann Arbor, MI 48109-2143, United States of America.

## Abstract

The magnetospheric cusps are important sites of the coupling of a magnetosphere with the solar wind. The combination of both ground- and space-based observations at Earth have enabled considerable progress to be made in understanding the terrestrial cusp and its role in the coupling of the magnetosphere to the solar wind via the polar magnetosphere. Voyager 2 fully explored Neptune's cusp in 1989 but highly inclined orbits of the Cassini spacecraft at Saturn present the most recent opportunity to repeatedly studying the polar magnetosphere of a rapidly rotating planet. In this paper we discuss observations made by Cassini during two passes through Saturn's southern polar magnetosphere. Our main findings are that i) Cassini directly encounters the southern polar cusp with evidence for the entry of magnetosheath plasma into the cusp via magnetopause reconnection, ii) magnetopause reconnection and entry of plasma into the cusp can occur over a range of solar wind conditions, and iii) double cusp morphologies are consistent with the position of the cusp oscillating in phase with Saturn's global magnetospheric periodicities.

## 1 Introduction

The magnetospheric cusps are an important site of plasma entry into the terrestrial magnetosphere and play a key role in the transfer of energy and momentum from the solar wind into the magnetosphere. Ground-based and in-situ observations at Earth have made much progress in the study of the cusp and the coupling of the solar wind-magnetosphere-ionosphere through the polar magnetosphere, see for example Smith and Lockwood (1996) and Cargill et al. (2005) and references therein for recent reviews. In-situ observations of the polar magnetospheres of the outer planets are restricted to some observations at Jupiter and Neptune. Voyager 2 fully explored Neptune's cusp in 1989 and the particle and field observations have been discussed thoroughly by Szabo et al. (1991) and Lepping et al. (1992). During the encounter of Ulysses with Jupiter in 1992, auroral hiss was observed during the inbound leg at lower latitudes (Stone et al., 1992) and data showed that Ulysses passed through open field lines on its inbound leg (Phillips et al., 1993), and polar cap field lines on its outbound leg (Bame et al., 1992; Cowley et al., 1993). High-latitude observations in Saturn's magnetosphere, made by the Cassini spacecraft in 2006-2009 present the most recent opportunity to study the polar magnetosphere of another rapidly rotating giant planet, and to examine the role of solar wind forcing in these magnetospheres. Most recently, Jasinski et al. (2015) presented a case of magnetosheath-like ion and electron distributions in Saturn's northern cusp, indicating reconnection with the solar wind at the magnetopause. Furthermore, the ion time-energy spectrograms showed evidence of stepped ion dispersions that are suggestive of bursty reconnection at the magnetopause.

Reconnection at the magnetopause is a fundamental process which opens the magnetosphere and allows solar wind plasma to enter the system; the plasma in the terrestrial magnetosphere is dominated by plasma of solar wind origin. The first observations of magnetosheath plasma in Earth's polar magnetosphere showed that fluxes of electrons and protons with magnetosheath-like distributions were observed just poleward of the last closed field line, near 75-79° magnetic invariant latitude, identified by a sharp decrease in the flux of >10 keV electrons as one moved onto open magnetic field lines (Heikkila and Winningham, 1971; Frank, 1971; Russell et al., 1971). Subsequent work, especially utilizing the multi-spacecraft capabilities of Cluster has shown the terrestrial cusp to be a broad complex region with a variety of interesting boundary layers, some of which form by the complex non-steady nature of dayside and high-latitude (lobe) reconnection (e.g. Cargill et al., 2005). The effects of magnetopause reconnection and its time variability can identified via latitude-energy and pitch angle-energy dispersions (e.g., Reiff et al., 1977; Burch et al., 1982; Lockwood and Smith, 1994). The orientation of the IMF plays a significant role in determining the motion of newly opened flux tubes at the dayside magnetopause (e.g., Cowley and Owen, 1989; Cooling et al., 2001) which contributes to the complex spatial morphology of the cusp, which can be separated into multiple entry regions (e.g., Wing et al., 2001; Zong et al., 2008; Pitout et al., 2009; Abe et al., 2011).

Identifying the signatures of magnetopause reconnection at Saturn has been the focus of a number of studies using in situ Cassini data. Lai et al. (2012) searched for evidence of flux transfer events by surveying 71 magnetopause crossings over a four-hour local time interval centered on 1200 Saturn local time and up to ~30º latitude, but didn't find evidence of local magnetopause reconnection. McAndrews et al. (2008) used two case studies to show evidence of reconnection on the dawn flank and also of lobe reconnection. Desroche et al. (2013) have shown that due to the combined effects of diamagnetic drift and flow shear, magnetopause reconnection is generally favored on the dusk flank and at higher latitudes away from the subsolar point. Evidence for reconnection at higher latitudes (>24º) was also presented by Badman et al. (2013) who also showed that bursty reconnection was present during compressions of Saturn's magnetosphere. In related work, Fukazawa et al. (2007) used a global magnetohydrodynamic simulation of Saturn's magnetosphere to show that the largest energy input into the polar cusp region was during northward ($B_z>0$) IMF and with a magnetopause reconnection site located northward of the sub solar point. These modeling studies therefore support the idea that magnetopause reconnection generally occurs away from the subsolar point and is more concentrated in the dusk sector.

At Earth, the orientation of the IMF is known to strongly affect the location of cusp auroral emissions (e.g., Wing et al., 2004 and references therein). Under southward IMF ($B_z<0$) the cusp auroral emission is located on or close to the main oval, and shifts poleward under northward IMF ($B_z>0$) when lobe reconnection takes place. The sign of the east-west component of the IMF ($B_y$) shifts the cusp emissions in magnetic local time. Bunce et al. (2005) studied the role of the IMF at Saturn and developed models of the flows and currents in the ionosphere produced by low latitude dayside and high latitude lobe reconnection. This modeling has shown that the sign of IMF $B_z$ affects the position of the cusp aurora, as it does at the Earth. Bunce et al. (2005) have also shown that the direct entry of magnetosheath electrons into Saturn's cusp would not be expected to produce measurable auroral emissions. However, pulsed reconnection at Saturn's magnetopause may produce field-aligned current systems that are of sufficient intensity to produce spot-like auroral emissions near the cusp.

Pallier and Prangé (2001) first identified high latitude auroral emissions at Jupiter that did not corotate and remained close to noon in magnetic local time and so were interpreted as the optical signature of the jovian cusp. Gérard et al. (2004) found similar features in Hubble Space Telescope (HST) images of Saturn's southern far ultraviolet aurorae where they noted the appearance of a bright (10-20 kR) spot located at approximately 15° co-latitude which was slightly poleward of the main auroral oval, somewhat distributed by ~1 hour of local-time either side of noon. Gérard et al. (2005) studied HST images of Saturn's southern auroral emissions in concert with upstream solar wind observations when Cassini was ~0.2 AU upstream of Saturn. Bright auroral spots were found at the onset of a period of minor compression in the solar wind. Gérard et al. (2005) interpreted auroral spots in the noon sector as the result of field-aligned currents produced by pulsed reconnection at the magnetopause (Bunce et al., 2005). Radioti et al. (2011) and Badman et al. (2013) have provided evidence for reconnection occurring at multiple locations on Saturn's magnetopause. Bunce et al. (2008) presented evidence for a crossing of auroral field lines connecting Cassini to Saturn's main auroral emission. During one period discussed in this study Cassini moved equatorward from the polar cap, through a population of magnetosheath-like plasma inside the magnetosphere and then into a region of hot electrons with evidence of field-aligned currents. Bunce et al. (2008) interpreted this region of magnetosheath-like plasma as particle entry in the cusp.

In this report we present two detailed case studies which discuss observations from the polar cap, mid-altitude cusp, and dayside boundary layers on two separate passes of Cassini at high invariant latitudes. Ion energy-pitch angle dispersions are presented that are interpreted as evidence for magnetopause reconnection driving particle entry in the cusp from different reconnection sites on the magnetopause. The entry in the cusp is often unsteady providing evidence that magnetopause reconnection is unsteady as well as occurring at different locations on the magnetopause during a given pass at high latitudes. It is argued that the position of the southern polar cusp periodically oscillates in a manner that is decoupled from motions driven by the upstream solar wind and is, therefore, probably related to global periodicities in Saturn's magnetosphere. Evidence is presented for boundary layers separating the cusp from the closed magnetosphere and for a hot electron population which is associated with Saturn's main auroral emission that is on closed magnetic field lines. The paper is organized into the following sections: in section two we describe some of the physics of rapidly rotating magnetospheres and discuss how this may modify the signature of the cusp compared to understanding based on the terrestrial magnetosphere. Instrumentation and data reduction is described in section three. The trajectory of Cassini during the two case studies is presented in section four. The case studies are presented in sections five and six. We conclude with a discussion in section seven.

## 2    Physics of the cusp at giant planets

The rapidly rotating magnetospheres that surround Jupiter and Saturn differ from the terrestrial magnetosphere in a number of important ways. These differences may affect interpretations of cusp behavior and properties that are based on understanding gained from the study of the terrestrial magnetosphere.

Most of the plasma in the saturnian magnetosphere originates from within the magnetospheric cavity due to mass-loading in the vicinity of the icy satellites, such as Enceladus and Dione, and in the E-ring torus (see Arridge et al. (2012) for a recent review). Ion populations in the magnetosphere vary greatly with radial distance, but are typically a mix of cold ($\lesssim$100 eV), warm (100 – 1000 eV) and hot (energetic) (>1 keV) populations. The electrons can similarly be divided into cold ($\lesssim$20 eV), warm (~100 eV) and hot (>500 eV) populations. The energetic populations are typically power law tails on the warm populations, such that the warm and hot populations can be described using Kappa distributions. Centrifugal forces are important and cold/warm heavy ions are centrifugally confined to the equator and form an equatorial plasma sheet in a similar manner to the equatorial confinement of iogenic plasma in the jovian magnetosphere (e.g., Hill and Michel, 1976). Lighter ions and electrons, and more energetic populations have much larger centrifugal scale heights than the cold/warm ions and are free to fill magnetospheric flux tubes to high latitudes (e.g., Sergis et al., 2011). However, because the heavy ions dominate the ion composition, polarization electric fields exist which pull the lighter species towards the equator in order to maintain charge quasi-neutrality (e.g., Maurice et al., 1997). As a consequence, the plasma at high latitudes should be dominated by hot electrons and energetic ions (light and heavy species) as these can overcome the field-aligned electrostatic potential, which is of the order of tens of volts at Saturn (Maurice et al., 1997).

The structure of the upstream solar wind at Saturn has been the focus of a number of studies (e.g., Jackman et al., 2004; Jackman et al., 2008; Jackman and Arridge, 2011).

Typically the upstream medium is organized into a pattern of corotating interaction regions (CIRs) which arrive quasi-periodically at Saturn once or twice per solar rotation period. These CIRs are separated by rarefaction regions where the solar wind has a relatively low dynamic pressure and the IMF has a very low field strength ≲0.1 nT. Inside a CIR the dynamic pressure and field strength are much higher and the clock angle of the field undergoes rapid changes in orientation (Jackman et al., 2004). Due to the large heliocentric distance of Saturn, the IMF is significantly wound-up with an average spiral angle of 86.75º (Jackman et al., 2008; Jackman and Arridge, 2011). Thus the average IMF has a dominant $B_Y$ component and newly opened field lines will tend to contract eastwards and westwards (in different hemispheres) due to the tension force on the newly open flux tubes (e.g., Cooling et al., 2001).

The rate of production of open flux, the reconnection voltage, has been estimated at Saturn using a "half-wave rectifier" function that produces mean reconnection voltages of 41.8 kV (Jackman and Arridge, 2011), but peak voltages of 100 – 400 kV in CIRs and 10 kV or less in rarefaction regions (Jackman et al., 2004). However, these estimates assume that the efficiency of dayside reconnection is the same as that at the Earth and the validity of this assumption has been the subject of some debate (e.g., Scurry and Russell, 1991; Masters et al., 2012; Masters, 2015). Whilst the dynamic pressure and field strength increases in CIRs – thus providing more favorable conditions for reconnection – the rapid oscillations in clock angle will modulate the reconnection rate and location on the magnetopause where reconnection can occur.

The scale of the jovian and saturnian magnetospheres also introduces further important differences. Saturn's equatorial radius is almost 10 times larger than that of Earth, and the subsolar magnetopause is around twice as far from the planet (in units of planetary radii) at Saturn compared to Earth, therefore the linear size of the magnetosphere is around a factor of 20 larger than the terrestrial system. The Alfvén wave travel time from the magnetopause to the ionosphere is of the order of an hour. A 1 keV proton on a newly reconnected field-line with a reconnection site near the subsolar point at Earth will take around 5 minutes to traverse the ~20 $R_E$ distance (1 $R_E$ = 6378 km) from the X-line to a spacecraft in the mid-altitude cusp. For a distance of 25 $R_S$ (1 $R_S$ = 60268 km) in the saturnian magnetosphere a 1 keV proton will take around an hour to travel from the reconnection site to the cusp.

The morphology of particle signatures in the cusp are dependent on the spacecraft speed, $v_S$, the convection speed of the plasma, $v_C$, and the effective speed of the open/closed field line boundary at the spacecraft, $v_B$, which is clearly related to the reconnection rate. In the Earth's magnetosphere at low altitudes $v_S$ is much larger than $v_B$ or $v_C$ and so the cusp is essentially at rest compared to the spacecraft speed. In the high-altitude cusp, however, the spacecraft speed is essentially irrelevant since $v_S$ is much smaller than $v_B$ or $v_C$. In the mid-altitude cusp $v_s$ can be similar to $v_b$ and $v_C$ thus providing a mixture of time-dependent signatures (Lockwood and Smith, 1994).

At Saturn, the speed of the open/closed boundary due to dayside reconnection can be estimated using published estimates of the dayside reconnection voltage (assuming no tail reconnection) and the magnetic flux through the polar cap, using the flux function from Cowley and Bunce (2003) with zonal internal field coefficients $g_n^0$ (to third order) from Cao et al. (2011). Since dayside reconnection increases the magnetic flux, Φ, through the polar cap thus increasing the size of the polar cap, θ. By taking derivatives of the flux function we can write the rate of change of θ as a function of the reconnection rate, $\mathrm{d}\Phi(\theta)/\mathrm{d}t$, and hence calculate the co-latitudinal speed of the boundary.

$$\frac{\mathrm{d}\Phi(\theta)}{\mathrm{d}t} = \frac{d}{d\theta}\left\{R_S^2 \sin^2\theta \left[g_1^0\left(\frac{R_S}{r}\right) + \frac{3}{2}g_2^0 \cos\theta \left(\frac{R_S}{r}\right)^2 + \frac{1}{2}g_3^0(5\cos^2\theta - 1)\left(\frac{R_S}{r}\right)^3\right]\right\}\frac{d\theta}{dt} \qquad (1)$$

Using a dayside reconnection voltage of 400 kV, the open/closed boundary will move equatorwards at ~0.006 °/hour, corresponding to a speed of 100 m s$^{-1}$ at a distance of 10 $R_S$. This small boundary speed compared to Earth is due to the much larger amount of magnetic flux through Saturn's polar cap compared to Earth.

The open/closed boundary is also expected to move due to oscillations in the magnetosphere. Despite the near axisymmetry of the saturnian internal magnetic field, magnetospheric periodicities have been noted since the Pioneer and Voyager flybys (see Carbary and Mitchell (2013) for a recent review). Recent studies using Cassini data have argued for the presence of rotating sheets of field-aligned currents (e.g., Southwood and Kivelson, 2007; Provan et al., 2009a, Andrews et al., 2010) which modulates the field configuration at high invariant latitudes, causing the cusp to spatially oscillate at a period close to that of the planet's rotation. Nichols et al. (2008) have used UV auroral images of Saturn's southern main auroral oval to show that the main oval oscillates at a period close to that of other magnetospheric periodicities (e.g., Provan et al., 2009b) with an amplitude of around 1º. Using the results of Nichols et al. (2008) we estimate a speed of ~120 km/s for the speed of the open/closed boundary at a distance of 10 $R_S$ (assuming it moves rigidly with the oscillation of the auroral oval). Hence, we expect the boundary motion to be mainly controlled by this oscillatory motion at high invariant latitudes.

In the events presented in this study the spacecraft is geometrically in the mid-altitude cusp (see the trajectories in Figure 1). The spacecraft speed relative to the planet is typically ~6 km s$^{-1}$. To estimate an upper limit for the solar wind-driven convection speed of the plasma we estimate the solar wind convection electric field, and assume that this maps into the magnetosphere with 100% efficiency to provide an upper limit to the convection speed. Using $v_{SW}$=400 km s$^{-1}$ and IMF $B_Z$=0.5 nT we obtain $E_{SW}$=0.2 mV m$^{-1}$ such that $v_C$, given by $E_{SW}/B$, is equal to 10 km s$^{-1}$ for a magnetospheric field strength of 20 nT. However, Saturn's magnetosphere is also rapidly rotating and the corotational convection electric field also plays a role. This is not important at Earth when calculating $v_c$. The ionosphere in the polar cap typically sub-corotates at a rate equal to one third of rigid corotation (Stallard et al., 2004) resulting in an azimuthal convection speed of 27 km s$^{-1}$ at 13 $R_S$ and 45º latitude. The spacecraft speed is therefore smaller than either solar wind-driven or sub-corotational flow. Hence, in the rest frame of the open/closed boundary, the satellite speed is comparable to, or smaller than that of the plasma convection speed and hence in the geometrical mid-altitude cusp at Saturn the observed particle signatures should have more in common with the high-altitude cusp signatures at Earth.

In comparing terrestrial observations of the cusp to the observations in this study at Saturn we must consider that a) plasma composition and energy spectra at high latitudes will not necessarily reflect that at low latitudes; b) dayside reconnection is more likely in CIRs but may well be bursty; c) the particle signatures have more in common with the high-altitude regime at Earth combined with low-altitude cusp effects due to the long transit times for a particle from the magnetopause to the spacecraft; d) the cusp might oscillate in position due to magnetospheric periodicities; e) the IMF spiral angle is large at Saturn providing a significant IMF $B_Y$ component which will affect the motion of newly opened field lines and possibly also the location of the reconnection site.

## 3      Instrumentation and data reduction

This study uses data from the Cassini magnetometer (Dougherty et al., 2004), Cassini Plasma Spectrometer (CAPS) (Young et al., 2004), Magnetospheric Imaging Instrument (MIMI) (Krimigis et al., 2004), and Radio and Plasma Wave Science (RPWS) (Gurnett et al., 2004).

Data from the magnetometer is taken from the fluxgate magnetometer at a cadence of 1s and is presented in spherical polar coordinates (Kronographic Radial-Theta-Phi – KRTP), based on the kronographic position of the spacecraft, where $e_r$ is along a vector from the planet to the spacecraft, $e_\theta$ points in the direction of increasing co-latitude, and $e_\varphi$ points azimuthally around Saturn in a prograde direction.

Plasma data come from the CAPS suite of instruments, specifically the Electron Spectrometer (ELS) and Ion Mass Spectrometer (IMS). ELS is a hemispherical top-hat electrostatic analyzer measuring electrons between 0.6 and 28750 eV/e with an energy resolution of 16.7%. Each of the eight anodes has an angular resolution of 20°×5.2° providing a 160°×5.2° instantaneous field-of-view, which is extended by rotating the instrument, which sweeps out ~200° of azimuth in ~4 minutes providing ~$2\pi$ sr total field-of-view. ELS captures data at a cadence of 2s which is sometimes down-sampled/averaged internally in the CAPS data processing unit before transmission to the ground. Pitch angle distributions are accumulated over a 4-minute azimuthal sweep of the instrument. Electron moments are calculated by integrating these pitch angle distributions to produce density and temperatures parallel and perpendicular to the field (Arridge et al., in preparation). Previous techniques (Lewis et al., 2008; Arridge et al., 2009) used numerical-integration of one-dimensional electron energy distributions with the assumption of isotropy in the spacecraft frame. Such moments are susceptible to anisotropies in the electron distribution and can produce unrealistic time-variations in density and temperature as the instrument samples different regions of velocity space. This new technique combines samples from azimuthal sweeps to generate 2D distributions in pitch-angle and energy thus removing such effects.

IMS is a hemispherical top-hat electrostatic analyzer measuring positive ions between 1 and 50 280 eV/q with an energy resolution of 16.7%. Each of the eight anodes has an angular resolution of 20°×8.3° providing a 160°×5.3° instantaneous field-of-view, which as for ELS, is also extended to almost ~$2\pi$ sr total field-of-view with azimuthal scanning. The highest time resolution data available from IMS is 4s. IMS also has a time-of-flight section to obtain energy-resolved mass per charge spectra with a mass resolution of 12.5%. Counts were scaled with energy-dependent efficiencies (H.T. Smith, private communication) for $H^+$ and $W^+$ (taken to be $O^+$) and with an average of $H_2^+$ and $He^{++}$ since they cannot be separated in IMS. Uncertainties in compositional ratios are based on counting statistics.

The MIMI Charge Energy Mass Spectrometer (CHEMS) instrument is used to provide energetic ion composition with an energy range of 3 – 220 keV/e, a mass per charge resolution of ~8% and a mass resolution of ~15%. The Low Energy Magnetospheric Measurements (LEMMS) sensor provides energetic ion and electron fluxes from 0.03 – 160 MeV for ions and 0.015 – 5 MeV for electrons. Plasma wave data are provided by the RPWS instrument which includes three nearly orthogonal electric field antennae, in order to detect AC electric fields between 1 Hz and 16 MHz. Calibrated high frequency (kilometric) emissions were produced using the method of Lamy et al. (2008).

## 4 Trajectory and data overview

Figure 1 shows the trajectory (figures 1a and 1c) of Cassini projected onto the noon-midnight meridional plane and its mapped ionospheric footprint (figures 1b and 1d) for Cassini revolutions (revs) 37 (figures 1a and 1b, 08 – 24 January 2007) and 38 (figures 1c and 1d, 24 January – 9 February 2007).

During rev 37, on 13 January 2007, Cassini crossed the tail plasma sheet and passed into the southern magnetotail lobe moving towards the dayside via the dawn flank. Cassini remained in the southern magnetic hemisphere until late on 17 January 2007. During the 16 January 2007 principal case study interval (indicated by the bold interval on the trajectory) Cassini was at high magnetic latitudes, magnetically-mapping to near the statistical UV auroral oval, and located between 1000 and 1200 SLT in the magnetic field region where we might expect to see the cusp. During rev 38, Cassini followed a trajectory through the magnetosphere that is very similar to rev 37, although at a somewhat larger radial distance in the cusp region. The field lines are more stretched corresponding to the lower upstream dynamic pressure during this event (see sections 5 and 6 for more details of the upstream conditions).

Figure 2 presents an overview of the data from rev 37 and 38. In both cases Cassini starts in a region where the plasma electron data is at the instrument noise level, the field strength is high and smoothly varying, and the energetic ion and electron fluxes are at the instrument noise level. We identify this region as the polar cap (indicated by the symbol PC). Following this interval Cassini enters a region with high field strength (although not as high as the polar cap and there are notable occasional depressions in the field strength) and large fluxes of low energy plasma electrons, consistent with magnetosheath plasma which is interpreted as magnetosheath particle in the cusp/boundary layers. These layers are alternately mixed with higher energy plasma consistent with closed field lines, before entering a region that we identify as the magnetosphere proper with higher fluxes of electrons and energetic particles on closed field lines.

In the rev 38 case (figure 1d-1e) we can see that a significant compression of the magnetosphere occurred around the middle of 02 February 2007 where Cassini enters the magnetosheath and also briefly the solar wind, thus providing an opportunity to compare and contrast the particle entry into the cusp and the particles in the adjacent magnetosheath. Even though the spacecraft is at a slightly larger radial distance during this orbit, entry into the magnetosheath and solar wind at such distances reflects an unusually high solar wind dynamic pressure. More details on the solar wind conditions are in sections 5.1 and 6.1. Towards the end of the interval Cassini has moved into the northern hemisphere in the afternoon local time sector and once again sees lower energy plasma, consistent with the cusp, before returning to the polar cap. We tentatively identify this lower energy plasma region as a mantle. In the rest of this paper we focus on the southern hemisphere observations at the beginning of these intervals.

# 5 Case study: 16 January 2007 (Rev 37)

## 5.1 Upstream conditions

Figure 3 presents the upstream conditions during the event as obtained from the MSWiM solar wind propagation model (Zieger and Hansen, 2008) which propagates solar wind observations from 1 AU to Saturn using a 1.5-D MHD model. These data are provided at a one-hour time resolution. Zieger and Hansen (2008) have comprehensively investigated the accuracy of these propagations and have shown, using the arrival times of solar wind shocks, that the temporal uncertainty is ±15 hours when Earth and Saturn are near apparent opposition (as they are during the events covered in this study). Hence, Figure 3 shows the solar wind propagations along with lagged time series to account for this uncertainty. A large solar wind disturbance arrived at Saturn on 06 January 2007 and the solar wind returned to a quiescent state on 14 January 2007 before another smaller enhancement occurred on 16 January 2007.

Figure 3 also shows the power of left-hand circularly polarized extraordinary-mode (X) emissions known as Saturn Kilometric Radiation (SKR), as measured by the RPWS instrument on Cassini, and integrated between 3 – 1000 kHz. The left-handed emissions originate from the southern hemisphere. Desch (1982) and Desch and Rucker (1983) have shown that SKR power is mainly controlled by the dynamic pressure of the solar wind. We have therefore used the SKR power to attempt to correctly lag the solar wind propagations. An increase in SKR power is found at 0600 UT on 16 January 2007 and an increase in solar wind dynamic pressure is found in the solar wind time series 14 hours before. Hence, we lag the solar wind propagations by +14 hours to align the increase in dynamic pressure with the increase in SKR power.

During the interval of the 16 January 2007 event the predicted dynamic pressure is 0.042±0.005 nPa and the IMF field strength 0.31±0.03 nT, which is mainly contained in the tangential component of the IMF (in radial-tangential-normal (RTN) coordinates). During this period the tangential direction in KSM coordinates is (0,-0.92,-0.38) hence a negative tangential component is oriented approximately duskward. The normal component in the propagations is essentially uncorrelated with observations, and the radial component cannot be propagated due to the 1.5-D nature of the MHD code. Using the model of Kanani et al. (2010), the average dynamic pressure during the event corresponds to a magnetopause subsolar distance of 19±3 $R_S$ which is a typical value (e.g., Achilleos et al., 2008). The quiet interval on 14 January 2007 corresponds to a subsolar magnetopause distance of 28±5 $R_S$. Hence, the arrival of the compression at 0600 UT on 16 January 2007 corresponds to a significant compression of Saturn's magnetosphere.

## 5.2   Overview and interpretation

Figure 4 presents an overview of the magnetic field, plasma/particle, and plasma wave observations during the 16 January 2007 event. The composition data were derived by summing counts in CHEMS, and in the IMS time-of-flight system, from the straight-through (ST) detector. To improve the counting statistics of the compositional data the counts were summed within the intervals identified in the CAPS/ELS data. Electron pitch angle distributions were analyzed during the course of this study but provide an ambiguous result due to the lack of full pitch angle coverage (the accessible pitch angle range is typically ~45º - 170º) in the downward (precipitating) direction, and to significant temporal variability in the data which aliases computed pitch angle distributions. Hence, electron pitch angle distributions are not presented for this event and will be the subject of future study.

At 0900 UT energetic particle (Figure 4b), plasma electron (Figure 4c) and plasma ion fluxes (Figure 4d) are at or near the noise level consistent with very low plasma densities. The electric field wave power shows considerable enhancement at low frequencies with a cut-off at ~110 Hz. Auroral hiss is commonly observed at high latitudes in Saturn's magnetosphere and generally consists of a whistler-mode emission below the electron cyclotron frequency, with a sharp cut-off at the electron plasma frequency (Gurnett et al., 2009). In this region the ELS data is at the instrumental noise level and so the electron density can be considered to have an upper limit of ~500 m$^{-3}$ (Arridge et al., 2009). The electron plasma frequency corresponding to this upper limit is ~200 Hz and the electron cyclotron frequency is higher, hence we attribute these emissions to auroral hiss. The CHEMS energetic ion composition indicates measurable fluxes of water group ions ($W^+$), $H_2^+$, and $He^{++}$. The latter is characteristic of solar wind plasma, but $W^+$ and $H_2^+$ are characteristic of closed magnetic field lines, where $W^+$ originates from neutral-plasma chemistry in the inner magnetosphere as a result of internal mass loading, and $H_2^+$ originates from Titan.

After 0955 UT, and until 1127 UT, fluxes of low energy ions and electrons are observed where the electron temperature (figure 4h) of 20 eV is consistent with a magnetosheath population. The electron fluxes are quite unsteady producing factor of ~four variations in the electron density. Figures 4e and 4f show the fraction of counts in CAPS/IMS and MIMI/CHEMS produced by various species. Each panel sums to 100% but the scales have been reduced to focus on the relative counts of species other than $H^+$. The plasma ion composition (figure 4e) appears to be devoid of heavy ions but the ratio of ions with mass per charge (m/q) of 2 amu/q (m/q=2) to $H^+$ counts is 3.7±0.3 %. $He^{++}$ and $H_2^+$ have an m/q of 2 amu/q and in the solar wind we would expect $He^{++}$ to dominate with around 4% $He^{++}$ whereas in the magnetosphere $H_2^+$ originates from Titan. Thomsen et al. (2010) have shown that the abundance of ions with m/q=2 relative to $H^+$ in the magnetosphere is around 20% inside 20 $R_S$ (in the equator) and around 5% beyond this distance. Hence, a ratio of ions with an m/q of 2 to $H^+$ of 20% is indicative of magnetospheric plasma whereas a value smaller than that indicates either solar wind plasma composition, or magnetospheric plasma at high latitudes where the $H_2^+$ is more concentrated towards the equatorial regions. The energetic ion composition (figure 4f) is $W^+/H^+$: 7±2 %, $H_2^+/H^+$: 5±2 %, $He^{++}$ 10±3 %. The relatively high $He^{++}$ fraction is suggestive of the presence of magnetosheath/solar wind plasma. In plasma ion data, a dispersion can be seen in the low energy edge of the ion distribution, which increases with time from 0955 to 1127 UT. The field strength (figure 4g) is ~20 nT, however its direction and stability is not consistent with an excursion into the magnetosheath which is usually characterized by very strong magnetic fluctuations.

The line at 1127 UT marks a compositional boundary based on the observations presented in Figure 4f. Between 1127 UT and 1151 UT there is no net field-aligned current (FAC) (no significant gradient in the azimuthal magnetic field component), the $W^+$ fraction increases in energetic ions and also the thermal plasma (0 to 0.10±0.08 % - not visible on the scale of this plot), and the plasma electrons exhibit some energization with a drop in density by a factor of ~5.

From 1151 to 1521 UT the plasma electrons have temperatures of 100 – 1000 eV and high fluxes of energetic electrons are observed. The energetic ion composition shows a significant population of $W^+$ and $H_2^+$, and in the warm plasma ions we see significant $W^+$ ($W^+/H^+$ of 0.2±0.1 %) and m/q=2 species (m/q=2 to $H^+$ of 8.8±0.7 %). In this region a decrease in the azimuthal component of the magnetic field can be seen, consistent with a layer of field-aligned current (e.g., Bunce et al., 2008). The intense plasma wave emissions observed under 50 Hz are consistent with shot noise from energetic electron impact on the electric field antenna (e.g., Zouganelis, 2008).

From 1521 to 1803 UT the electron temperature drops to values consistent with that observed from 1127 and 1151 UT and the energetic electrons drop to close to near-background levels (a similar region is also observed from 1901 to 2050 UT with a short excursion back into the region with hot electrons). The plasma composition maintains a large water group ion component but the $He^{++}/H^+$ ratio increases by a factor of two at the beginning of the interval, moving to a factor of 20 after 1728 UT. The intense interval from 1728 to 1803 UT has a significant diamagnetic depression and contains a mixture of thermal ion populations.

To determine the flow direction of the ions Figure 5 shows ion counts plotted as a function of look direction around the spacecraft. This also enables us to identify what directions around the spacecraft are not visible to IMS. These data are presented as a polar projection of OAS coordinates which is a spacecraft-centered frame where **S** is a vector from the spacecraft to Saturn, **O** is a vector which is obtained from **S**×(**Ω**×**S**) and **A** is a vector along **S**×**O** and completes the right-handed set. In this projection the polar angle $\theta_{OAS}$ is the angle between a vector and **S** such that $\theta_{OAS}=0°$ represents a direction towards Saturn from Cassini (the center of each plot), and 90° is perpendicular to the Cassini-Saturn line (the inner circle on each plot). The outer circle on each plot is from the direction diametrically opposite to Saturn. Ion counts in the inner circle are coming from "in front" of Cassini, and between the outer and inner circles come from "behind" Cassini. The angle around **S** is identified as an azimuthal angle where counts from the left (right) half have a component in the corotational (anti-corotational) direction, from the upper (lower) half are coming from "above" ("below").

In Figures 5a, 5c and 5e ions with an energy/charge of 724.1 eV/q are found near $\theta_{OAS}≈84°-100°$ and $\varphi_{OAS}≈340°-348°$ which indicates ions flowing anti-sunward, duskward and in downward, with a vector approximately (-0.48, 0.42, -0.77) in KSM coordinates, thus these ions are flowing poleward with a significant component in the direction of azimuthal convection. The energy of these ions indicates a speed of 370 km/s but this is an upper limit to the speed since it assumes the ions are cold. In reality some of this energy is due to thermal motions of the ions. This calculation assumes $H^+$ which is consistent with the measured composition. In Figures 5b and 5d we find hot ions flowing approximately anti-parallel to the magnetic field but where some of the ions (at angles between the 180° pitch angle direction and $\theta_{OAS}≈90°$) are obscured by parts of the spacecraft bus (e.g., Young et al., 2004). In Figure 5f we find higher energy ions also moving in the poleward and duskward direction.

Returning to figure 4, from 1803 to 1901 UT and 2050 UT onwards the spacecraft once again encounters a region of hot electrons and magnetospheric plasma composition.

The region between 0955 and 1127 UT is interpreted as magnetosheath particle entry into the cusp due to the plasma composition, magnetosheath-like electron distributions and low energetic electron fluxes. The ion dispersion is such that the more energetic ions are found at the equatorward edge of this region similar to a normal-sense ion dispersion found in the terrestrial cusp (e.g., Reiff et al., 1977). This sense of dispersion is characteristic of magnetopause reconnection equatorward of the spacecraft. Thus we argue that Cassini is in the cusp on open field lines and observing magnetosheath plasma that has entered previously entered the magnetosphere via dayside reconnection. The bursty electron fluxes observed in the cusp may be the signature of unsteady magnetopause reconnection. Because of the incomplete pitch angle coverage of CAPS we only have observations of ions and electrons to a pitch angle of ~45º and so we do not directly see precipitating (planetward-moving ions) and only directly see ions that have mirrored at low altitudes and are moving anti-planetward. It is not clear if there is ongoing injection of plasma from the magnetosheath, hence part of this region might be properly referred to as the start of the "mantle" (e.g., Rosenbauer et al., 1975). However, the ion energy-pitch angle dispersions show significant planetward fluxes at pitch angles of ~45º hence we retain the identification of this region as the cusp but recognize that we are also probably observing the start of the mantle.

The extended region from 1155 to 1521 UT contains hot electrons and a magnetospheric ion composition and so could conceivably be located on closed field lines equatorward of the cusp. This is also the region where we infer the presence of upward FAC connecting to Saturn's main auroral emission (Bunce et al., 2008). The boundary region from 1127 to 1155 UT separates the cusp from the auroral field lines, appears to correspond with zero FAC, and contains plasma with a composition intermediate between the cusp and the closed region. The increasing fraction of water group ions suggests this layer is on closed field lines but the solar wind composition of thermal ions (with a mass per charge of two) and presence of $He^{++}$ seen in the energetic particle composition suggests some mixing of solar wind and magnetospheric plasma, potentially via chaotic energetic ion trajectories. However, from the modified thermal electron populations and magnetic field rotations we interpret this region as a boundary layer, possibly representing the high latitude extension, or low-altitude projection of the low-latitude boundary layer. This layer may be on closed field lines although the lack of pitch angle coverage does not allow us to search for bidirectional electron beams that are characteristic of closed field lines.

The region before 0955 UT is interpreted as the polar cap on open magnetic field lines due to the i) very low plasma and energetic particle fluxes ii) the observation of auroral hiss, and iii) it's location poleward of the cusp. Although the energetic ion composition indicates the presence of magnetospheric ion populations, the gyroradius for a 40 keV $W^+$ ion is around 0.1 $R_S$ and so finite gyroradius effects might allow W+ ions to access to the open field region. Alternatively these ions could have a pitch angle near 90º and thus are slowly moving along the field line.

Another boundary layer is identified from 1521 to 1803 UT. The presence of magnetosheath-like electron distributions might suggest a re-entry into the cusp, supported by the increased amount of $He^{++}$. However, the large $W^+$ component and the fact that the electrons are accelerated compared to the magnetosheath suggests that this is a re-entry into the boundary layer that was observed from 1127 – 1155 UT. Interestingly, the $W^+$ fraction is typically larger in the boundary layer than in the adjacent auroral region, which might be produced by centrifugal confinement of heavy ion plasma in the closed-field region, but viscous mixing of magnetospheric and solar wind plasma in the boundary layer at lower latitudes combined with lower azimuthal velocities (due to viscous mixing) resulting in less centrifugal confinement. The two-population ion distribution found between 1728 and 1803 UT can be interpreted in two ways: i) poleward convection due to newly open field lines and azimuthal convection associated with IMF $B_Y$ and partial corotation, or ii) azimuthal motion due to partial corotation of a boundary layer. From the width of the distribution in energy and angle the ion population appears to be warm suggesting magnetospheric ions rather than magnetosheath ions. Also taking into consideration the high number of $W^+$ counts we interpret this as a boundary layer where the plasma has become mixed with water group ions from the magnetosphere that are sub-corotating.

In summary, the spacecraft starts in the polar cap and moves through a region with magnetosheath-like plasma that we identify as particle entry into the cusp due to dayside reconnection. At the equatorward boundary of the cusp we see a boundary layer that is conceivably on closed field lines, before entering a region with hot electrons probably mapping to the location of Saturn's main auroral emission, but which do not have sufficient thermal energy flux to directly produce auroral emission (Bunce et al., 2008). Cassini then alternates between this region and the boundary layer region before remaining on magnetospheric field lines at the end of the event.

## 5.3   Ion dispersions and dayside reconnection: estimating the distance to the reconnection site

At Earth the energy-latitude dispersions are produced by the differing time-of-flight of ions, injected at the magnetopause during magnetic reconnection. These ions have a field-aligned component of motion but are also on flux tubes that are moving in a global sense under the influence of solar-wind driven poleward convection (e.g., Reiff et al. 1977). Due to rapid planetary rotation at Saturn some azimuthal dispersion might also be present due to partial corotation on open field lines combined with azimuthal convection associated with IMF $B_Y$. The observation of an energy-time dispersion in figure 4 suggests that reconnection at Saturn's magnetopause is the cause of the particle injection. To estimate the field-aligned distance to the reconnection site we use the observed energy-pitch angle dispersions.

Burch et al. (1982) demonstrated that due to magnetic mirroring and ions of various energies having differing times-of-flight along the field line from a reconnection site, the ions should exhibit an energy-pitch angle dispersion in the cusp. Electrons are not expected to exhibit such a dispersion due to pitch angle scattering of the electrons during their transit from the magnetopause to the cusp. Equation 2 shows the ion energy cut-off as a function of pitch angle and time since reconnection

$$E(\alpha_0, t) = \frac{m_i}{2t^2} \left\{ \int_{s_i}^{s_o} ds \left[ 1 - \frac{B(s)}{B_o} \sin^2 \alpha_0 \right]^{-\frac{1}{2}} \right\}^2 \qquad (2)$$

where $m_i$ is the ion mass, $t$ is the time since reconnection, $s$ is position along a field line and $ds$ is an element along that field line, $B_0$ and $\alpha_0$ are the field strength and ion pitch angle at the observation point and $B(s)$ is the field strength along the field line.

Figure 6 shows measured ion energy-pitch angle dispersions in the cusp. We also analyze the ion observations in the boundary layer between 1521 and 1803 UT. The solid points show the measured low energy ion cut-offs, extracted by searching for where the ion flux drops below a signal-to-noise ratio of 4. The solid curves are a model fit of these cut-offs to equation 2 where both the distance to the reconnection site $D=s_o-s_i$ and the transit time $t$ are free parameters (the observation point, $s_o$, is known from the spacecraft position). The model was fitted using non-linear least squares using the Levenberg-Marquardt algorithm (Markwardt, 2009, and references therein) to minimize the difference between the model and the observed low-energy ion cut-offs.

The results of this analysis are presented in table 1. During the first particle injection in the cusp the mean distance is 50±20 $R_S$ and is inconsistent with reconnection near the sub-solar point but perhaps consistent with reconnection towards the flanks of the magnetosphere. In the boundary layer, ion energy-pitch angle dispersions are still observed suggesting mixing in the boundary layer due to reconnection. Hence, this boundary layer is possibly on open field lines. The mean distance to the reconnection site in this boundary layer is significantly smaller at 24±2 $R_S$, more consistent with a subsolar reconnection site, although the magnitude of the azimuthal field component indicates that the reconnection site is probably displaced in azimuth. These results demonstrate that the magnetosheath particles can access high latitudes due to magnetopause reconnection, possibly at more equatorial regions.

## 5.4 Periodic encounters with the cusp/boundary layer

From Figure 4 it can be seen that the cusp and boundary layer are encountered quasi-periodically. Such a morphology is also seen in the terrestrial cusp and has been interpreted as a spatial structure and as the consequence of a strong $B_y$ component of the IMF (Wing et al., 2001; Pitout et al., 2009; Abe et al., 2011) or as a temporal effect when the magnetosphere oscillates causing the location of the cusp to also oscillate (Zong et al., 2008). In support of the former interpretation for these data at Saturn, the nominal spiral angle of the IMF at Saturn's location is 87° (Jackman and Arridge, 2011) producing a nominally large $B_y$ component.

However, Cassini observations of the Saturn system have shown the magnetosphere to oscillate with a period of ~10.7 hours, close to that of Saturn's planetary rotation period, producing oscillations in magnetic fields, plasma, energetic particles, energetic neutral atoms, and associated radio emissions (e.g., Carbary and Mitchell, 2013). Nichols et al. (2008) reported observations and modeling showing that Saturn's southern auroral oval oscillates in position, with an amplitude of several degrees, in phase with these magnetospheric oscillations. With the open/closed field line boundary located at the poleward edge of the auroral oval the cusp position should therefore also oscillate at this period. Nichols et al. (2008) presented fits to the auroral data thus providing a time-dependent model for the location of the auroral oval that can be used to infer the location of the open/closed field line boundary relative to the spacecraft.

Using the fits from Nichols et al. (2008) we transform the ionospheric footprint of Cassini into the frame of the moving auroral oval, effectively providing a re-mapped invariant latitude and magnetic local time which takes into account the current systems associated with the periodic phenomena. Figure 7 shows this re-mapped invariant colatitude as a function of time along with electron data from CAPS/ELS for the January 2007 event. The blue curves in the bottom panel show the poleward and equatorward edges of the auroral oval from Carbary (2012) at the local time of the spacecraft. During this period Hubble Space Telescope images of the aurora are available for 0531-0541 UT on 16 January and 0321-0330 UT on 17 January (Bunce et al., 2008). The observations show that the colatitude of the auroral oval fell from ~15º to ~10º between the two images. Using an average of these two colatitudes we adjust the Carbary (2012) oval to better match the observed oval position – although we recognize that the oval position is contracting during this period we do not attempt to model this time-dependence. At the start of the interval the spacecraft is poleward of the auroral oval (the re-mapped invariant co-latitude is below the model auroral oval co-latitude) but approaches the oval as the spacecraft enters the cusp. The spacecraft moves through auroral oval field lines (see also Bunce et al., 2008) and equatorward of the oval before undergoing a poleward motion approaching, but not crossing the oval field lines again when the boundary layer is encountered. At the end of the interval the spacecraft is equatorward of the oval and oscillations of the oval are not of sufficient amplitude to allow the spacecraft to re-encounter the open field lines. The multiple encounters with the boundary layers between 1600 and 2000 UT on 16 January might be due to time-dependence of the auroral oval position, possibly produced via solar wind pressure variations.

## 5.5 Summary

During the interval studied Cassini starts in the polar cap before passing through magnetosheath particle entry into the cusp that is associated with magnetopause reconnection occurring at a reconnection site 50±20 $R_S$ from the spacecraft, thus somewhere on the flanks of the magnetopause. Cassini then passes through auroral field-lines before the oscillation of Saturn's high latitude magnetosphere moves Cassini closer to the open/closed field line boundary and thus onto boundary layer plasma that sits between the cusp and the auroral field lines. Eventually the high latitude magnetosphere rocks back moving Cassini back onto auroral field lines before passing through onto closed magnetospheric field lines. The final boundary layer encounter is interpreted as a dynamical event. From auroral imaging we know the oval is contracting at this time and hence this would tend to move Cassini in the opposite direction to that which is observed. Hence, we argue that dynamic pressure variations are most likely responsible for this behavior.

# 6 Case study: 01/02 February 2007 (Rev 38)

## 6.1 Upstream conditions

Figure 8 presents the propagated upstream conditions during the event in the same format as figure 2. Before the cusp passage Saturn was in a solar wind rarefaction region with a very low solar wind dynamic pressure of 0.0014±0.0003 nPa corresponding to a subsolar magnetopause distance of 38±6 $R_S$. Rather than using increases in SKR power to identify the correct lag for the solar wind time series we use the magnetopause crossing observed at 1126 UT on 02 February 2007 (see Figure 11 and section 6.2) in the in situ fields and particles data. The solar wind time series was lagged by -16 hours to match the increase in dynamic pressure with this magnetopause crossing. From this lagged time series we can see that a solar wind forward shock arrived at Saturn on 31 January, and a large increase in solar wind dynamic pressure occurred around 1200 UT on 02 February 2007. During the first part of the February 2007 event, as shown by the grey bars, the solar wind was at a lower dynamic pressure state of 0.0108±0.0006 nPa, field strength 0.113±0.007 nT, and tangential IMF component of -0.04±0.03 nT. Towards the end of the interval the dynamic pressure increased to 0.0395±0.0003 nPa, but the field strength decreased to 0.0647±0.0002 nT and the tangential component dropped to zero. The tangential direction in KSM coordinates is (0,-0.92,-0.38) hence, the field was very weakly duskward at the beginning of the event, and rotated to an orientation almost entirely in the X-Z plane of KSM towards the end of the interval. At the beginning of the event the subsolar magnetopause distance, calculated from the dynamic pressure using the model of Kanani et al. (2010), was 25±4 $R_S$, dropping to 19±3 $R_S$ at the end of the interval when the magnetopause was crossed. We note that the SKR power, particularly the low frequency power, increases at the arrival of the forward shock on 31 January, and also during the event.

## 6.2 Overview and interpretation

The February 2007 event follows a similar but somewhat more straightforward morphology to the January 2007 event. Figure 9 presents the observations in the same format as Figure 4. Cassini starts the interval in a region with plasma ion and electron and energetic electrons fluxes near or at the noise level, auroral hiss is observed in the electric field data. From 1533 to 1819 UT on 01 February, magnetosheath-like electron distributions are observed but with energetic electrons at the noise level. The energetic ion composition in this region has a large $W^+$ and $H_2^+$ component but no $He^{++}$ within error. The plasma composition shows no evidence of $W^+$ but a small fraction (~2%) of m/q=2 amu/q ions (m/q=2 to $H^+$ is 2.1±0.3%). A narrow layer is observed from 1819 to 1851 UT with an increased energetic $W^+$ fraction and increased plasma m/q=2 amu/q fraction. Energized electrons are observed in ELS and an increase in flux of energetic electrons is recorded by LEMMS. The magnetometer data shows no evidence of field-aligned currents.

From 1851 – 2341 UT hot electrons and substantial fluxes of energetic electrons are found. The magnetometer data shows some evidence for FAC in a brief interval immediately after 1851 UT where rotations in $B_\varphi$ show evidence for an upward current layer is found, and a more distributed apparent downward layer towards the end of this region, but in light of the cusp motion identified in Figure 4 this downward layer is more likely to originate from the spacecraft moving back through the upward current layer in the opposite direction. The plasma composition is dominated by $H^+$ with m/q=2 to $H^+$ ratio of 5.7±0.9%, although the energetic ion composition is consistent with magnetospheric plasma with $W^+/H^+$ of 29±1%, $H_2^+/H^+$ of 14.4±0.8% and $He^{++}/H^+$ of 2.4±0.3%.

From 2341 UT on 01 February – 0005 UT on 02 February we see a narrow layer with energized sheath-like electron distributions, a drastically reduced $W^+$ fraction, and an increase in $He^{++}$. The magnetometer data shows little evidence for FACs.

The region between 0005 and 0246 UT contains substantial fluxes of magnetosheath-like electron distributions and a large fraction of $He^{++}$ consistent with solar wind plasma, although substantial fluxes of magnetospheric ions ($W^+$ and $H_2^+$) are also found in this region. The plasma ions show evidence for an energy dispersion with the most energetic ions and no low energy ions found on entry into this region. A diamagnetic depression is found in this region.

From 0246 to 1126 UT the ion composition is magnetospheric, with 0.4±0.1% $W^+/H^+$ and 15.4±0.7% m/q=2 to $H^+$ from IMS and $W^+/H^+$ of 23.7±0.5%, $H_2^+/H^+$ of 17.7±0.4% and $He^{++}/H^+$ of 1.6±0.1% from CHEMS. Hot electrons and large fluxes of energetic electrons are observed and some evidence for upward FAC, equatorward of the cusp, are found in the magnetometer data.

Finally, from 1126 UT to the end of the interval the magnetic field strength is low and its orientation is highly variable. Large fluxes of magnetosheath-like electron distributions are observed and the plasma composition is largely solar wind, but with some added energetic ions. From IMS the composition was found to be $W^+/H^+$ 0.020±0.01% and m/q=2 to $H^+$ of 3.4±0.2%, and from CHEMS $W^+/H^+$ 10±2%, $H_2^+/H^+$ 11±2%, $He^{++}/H^+$ 7±1%.

Similar to the January 2007 interval we interpret the region before 1533 UT as the polar cap, after which Cassini crosses into a region with magnetosheath-like electron distributions and a mix of magnetospheric and solar wind plasma which we interpret as the cusp. The magnetospheric plasma that we find in the cusp may originate from energetic particles still on newly opened field lines and which are draining out of the magnetosphere, or may be the result of finite gyroradius effects, or ions with a quasi-perpendicular pitch angle and which are trapped in the cusp (e.g., Zhou et al., 2006). The narrow region from 1819 to 1851 UT with energized magnetosheath electrons and an increased population of magnetospheric ions is interpreted as another boundary layer, following the entry of Cassini onto closed field lines with a magnetospheric ion composition and hot electrons from 1851 to 2341 UT. We interpret the narrow layer from 2341 to 0005 UT as another boundary layer before entering the cusp again from 0005 to 0246 UT where we find significant fluxes of $He^{++}$ and magnetosheath-like electron distributions. In contrast to the 16 January 2007 event the energetic electron fluxes remain almost constant across this entry into the cusp suggesting a slightly different configuration to the February 2007 event. No boundary layer is found after exiting the cusp onto closed field lines. To support the identification of a closed field-line region, we note the magnetospheric ion composition, hot electrons, and very small fraction of $He^{++}$. The region after 1126 UT is the magnetosheath with a significant rotation in the magnetic field at 1126 UT. The magnetic field in the magnetosheath is highly variable in direction and strength, the magnetosheath-like electron distributions and $He^{++}$ is once again observed with a solar wind-like (~4%) fraction. This interval provides an opportunity to directly compare the magnetosheath with the cusp signature: for example the similarity in electron distributions near 0100 UT and 1200 UT on 02 February 2007.

In summary, Cassini starts in the polar cap, passes through the cusp on open field lines and then onto closed magnetospheric field lines with a boundary layer separating the cusp from the closed field lines. After approximately five hours Cassini then passes back through a boundary layer and into the cusp, then back onto closed field lines with no boundary layer visible in the data at this transition. Approximately 10.5 hours later Cassini crosses the magnetopause and enters the magnetosheath as a result of the solar wind compression which arrives at Saturn around 1200 UT on 2 February.

## 6.3 Estimating the distance to the reconnection site

Similar to the 16 January 2007 event, the distance to the reconnection site was estimated using the observed energy-pitch angle dispersions. Figure 12 and Table 2 shows the results of this analysis. Apart from the dispersions analyzed in the boundary layer in the 16 January 2007 interval, the reconnection site distances are comparable. For the two periods in the cusp in the February 2007 event the calculated distance to the reconnection site is the same, within the uncertainties, between the two cusp passages.

## 6.4 Periodic encounters with the cusp/boundary layer

Similar to the January 2007 event the various layers presented in section 6.2 are observed twice. However in this case the oscillations move Cassini properly into the cusp rather than merely entering a boundary layer. From Figure 8 we can see the remapped ionospheric footprint of Cassini with respect to a statistical auroral oval. The more expanded state of the magnetosphere during the February 2007 changes the mapping, resulting in oscillations whose latitudinal amplitude is smaller than the 16 January 2007 event but also allowing for more oscillations during the interval.

Figure 13 shows the cusp crossing data compared with the remapped invariant latitude in the same format as Figure 9. No images of the aurora are available for this interval and so we have adjusted the position of the statistical oval to match the first cusp entry by simply shifting the statistical poleward by 4º. The second cusp passage is not accurately reproduced by this remapping, however the upstream conditions are more variable during this interval with a forward shock reaching Saturn on 31 January and the increase in solar wind dynamic pressure on 02 February. Hence, the auroral oval and position of the open/closed boundary may be shifted due to the influence of the solar wind, possibly via on-going magnetopause reconnection causing equatorward expansion of the auroral oval. We also note the gradual increase in solar wind dynamic pressure around the time of the second cusp encounter. However, in general the oscillatory nature of the cusp/boundary layer encounters on both the January and February 2007 events are adequately explained by oscillations of the cusp position produced by global magnetospheric oscillations.

# 7    Discussion

## 7.1    Conclusions

In this paper we have discussed observations made by the Cassini orbiter during two passes through Saturn's polar magnetosphere. We have shown that i) Cassini directly encounters Saturn's southern polar cusp with evidence for the injection of magnetosheath plasma into the cusp via magnetopause reconnection, ii) the injection of magnetosheath-like plasma is variable suggesting that magnetopause reconnection is bursty, iii) the precipitating plasma can originate from a variety of locations on the magnetopause, iv) magnetopause reconnection and injection of plasma into the cusp can occur under a range of solar wind dynamic pressures, v) boundary layers separate the cusp from field lines with auroral electrons which map to Saturn's main auroral emission, vi) the position of the cusp oscillates in phase with Saturn's global magnetospheric periodicities.

In both case studies presented in this paper Cassini moves from the polar cap, passing through the cusp, and onto field lines with hot electrons (~1-10 keV) that map to Saturn's main auroral emission, although these electrons do not have the required energy flux to produce the main emission without further acceleration (Bunce et al., 2008). With only one exception, a boundary layer separates the cusp from the region mapping to the main auroral emission. This is consistent with Jinks et al. (2014) who found that the polar cap boundary (where one might expect to find particle injections in the cusp) is displaced from the upward FACs. In the cusp, significant fluxes of magnetosheath-like electron distributions (~20 eV) and ions (~100 – 1000 eV/q) with a solar wind-like fraction of He$^{++}$ were observed showing that the cusp is can be filled with plasma of solar wind origin.

Ion energy-pitch angle dispersions were used as evidence for magnetopause reconnection allowing entry of solar wind plasma into the magnetosphere, rather than the plasma simply gaining access to the magnetosphere through diffusive processes or directly entering via a weak field region in the cusp. Some evidence for energy-latitude dispersions in the ion data also supports this reconnection picture but these were not analyzed in order to attempt to estimate the distance to the reconnection site. This was not attempted due to large uncertainties in estimating the speed of the spacecraft relative to the open/closed boundary, the speed of poleward convection of open field lines, and the speed of azimuthal magnetospheric convection due to sub-corotation of the polar cap ionosphere. However, the observations are qualitatively consistent with expectations for the low energy ion cut-off energy falling with distance from the closed field region. The solar wind conditions were different between the two events showing that the cusp at Saturn is active under a range of upstream conditions and not purely during periods of strong magnetospheric compression and hence strong solar wind driving.

The composition in the boundary layers consisted of a mixture of magnetospheric and solar wind plasma showing the presence of mixing of the two populations. In one case a two-component ion population was found. The limited pitch angle coverage does not allow us to identify whether these boundary layers are on open or closed field lines but in one case where the two-component population was found one population exhibited an energy-pitch angle dispersion and so it was argued that at least this region of the boundary layer could be on open field lines.

In the 16 January 2007 event evidence for reconnection at two different locations on the magnetopause was presented, showing that the position of the magnetopause reconnection site can vary on relatively short timescales of several hours. Variability in the precipitating electron fluxes was used to infer the presence of bursty/unsteady reconnection at a given reconnection site. However, both examples of the southern cusp presented here are more quiescent that the stepped ion dispersions discussed by Jasinski et al. (2015).

During each pass the southern cusp and associated boundary layers were encountered twice, similar to double/triple cusp morphology observed in the terrestrial magnetosphere. Oscillations in the position of Saturn's auroral oval were used to provide evidence that this double cusp morphology is the result of global magnetospheric periodicities, under the assumption that the cusp was located at the poleward edge of the auroral oval. Figure 12 presents a summary of this argument showing the oscillation of the ionospheric footprint with respect to a statistical auroral oval (Carbary, 2012). In both cases the footprint generally moves equatorward but undergoes reversals where the footprint moves briefly poleward, thus providing an opportunity to re-encounter the cusp/boundary layer.

## 7.2   Implications and further work

Gurnett et al. (2010) reported the presence of a "plasmapause-like" boundary at high latitudes in Saturn's magnetosphere similar to the density gradient found between the auroral field lines and the polar cap in this study. Such a density gradient might be found on closed field lines where the centrifugal confinement of heavy ions to the equatorial regions reduces the plasma density to very low values at high latitudes. Also field lines at large *L* are very long and as a result the transit times of particles along the field becomes very long, comparable to the azimuthal convection time around the planet for the equatorial plasma. The region identified as the polar cap in this study was argued to be on open field lines based on its location poleward of the cusp and the very low fluxes of plasma and energetic particles. Hence, we argue that Cassini has sampled magnetically open field lines and not merely field lines that are still closed but where the plasma is centrifugally confined to the equator. Naturally this does not imply that other local times, or periods where the cusp region is not magnetically open, that such a boundary might not exist on closed field lines.

The cusp was found to be active during the two passes through Saturn's polar magnetosphere under a range of upstream conditions. A number of studies have argued that magnetopause reconnection should be a low efficiency process at Saturn (e.g., Scurry and Russell, 1991; Masters et al., 2012; Masters, 2015). The results in this study show that Saturn's cusp is active under a range of upstream conditions. Since the cusp maps to a very large area on the magnetopause these results show that reconnection can readily occur somewhere on the magnetopause under a range of upstream conditions, supporting the findings of Desroche et al. (2013) and Fuselier et al. (2014).

Bunce et al. (2008) used the January 2007 observations reported here to argue that Saturn's main auroral emission maps to the open/closed field line boundary at the location of a velocity shear between open and closed magnetic field lines. In this study we have identified the presence of a boundary layer lying between the open polar cusp and the auroral field lines. It is conceivable that this boundary layer is located on closed field lines and hence is an internal boundary layer. Hence, the location of the velocity shear responsible for the FAC that drive Saturn's main auroral emission might lie at the boundary between the magnetosphere at the boundary layer, thus placing the auroral field lines on closed flux. This is not incompatible with the findings of Bunce et al. (2008) since we might expect a velocity shear to exist between the boundary layer and the magnetosphere proper. Further study of the location of these FAC in relation to boundary layers is required to clarify this aspect of the generation of Saturn's main auroral emission.

The identification of periodicities in the location of the southern cusp show that studies inferring the presence and direction of FAC at high latitudes, and studies inferring the presence of lobe/near-subsolar reconnection from the sense of ion energy-latitude dispersions must account for the oscillations of the polar magnetosphere. We have established oscillations of the southern polar magnetosphere but further study is required to establish this for the northern polar magnetosphere (e.g., Bunce et al., 2014; Jinks et al., 2014).

General statements on the physics of the cusp in rapidly rotating giant planet magnetospheres were made in section two. The findings in this report have implications for the study of the jovian high latitude magnetosphere with Juno. Further study and modeling of the cusp in the jovian magnetosphere will be required in order to analyze Juno observations.

A detailed survey of this dataset including cusp encounters in the northern and southern hemispheres will be reported in a future paper (Jasinski et al., in preparation).

## Acknowledgements

All the data used in this paper can be obtained from NASA's Planetary Data System at http://pds.jpl.nasa.gov. We thank Lin Gilbert and Gethyn Lewis for data processing and software support for CAPS at MSSL/UCL, and Steve Kellock, Leah Nani-Alconcel and Peter Slootweg at Imperial College for MAG data processing, and engineers at University of Iowa and LESIA for RPWS data processing. CSA thanks Frank Crary, Tom Hill, Jared Leisner, Hazel McAndrews, Don Mitchell, Mark Shappirio, Todd Smith and Rob Wilson for useful comments and discussions. CSA was supported in this work by an STFC Postdoctoral Fellowship and Royal Society University Research Fellowship, AJC was supported by the STFC rolling grant to MSSL/UCL, SWHC/EJB by an STFC consolidated grant ST/K001000/1, MKD by a PPARC Senior Fellowship, PZ and LL were supported by CNES. Cassini CAPS and MAG operations at MSSL and Imperial College were supported by STFC and ESA.

situ Ulysses/URAP radio measurements in the solar wind, J. Geophys. Res. 113, A08111, doi:10.1029/2007JA012979.

# Tables

Table 1: Field-aligned distance to the reconnection site during the 16 January 2007 event as estimated from observed ion energy-pitch angle dispersions and a fit to Equation 1.

| Date | D [$R_S$] | T [hours] |
|---|---|---|
| 2007-01-16 10:24:19 | 70±60 | 10±7 |
| 2007-01-16 10:57:55 | 50±30 | 5±2 |
| 2007-01-16 11:04:51 | 50±40 | 5±3 |
| 2007-01-16 11:25:07 | 40±20 | 3±1 |
| 2007-01-16 11:32:03 | 80±60 | 5±4 |
| 2007-01-16 11:38:59 | 40±20 | 3±1 |
| **Average** | **50±20** | **5±2** |
| | | |
| 2007-01-16 17:24:35 | 16±3 | 1.1±0.2 |
| 2007-01-16 17:38:27 | 40±10 | 3.0±0.8 |
| 2007-01-16 17:44:51 | 24±6 | 2.1±0.5 |
| 2007-01-16 17:51:47 | 21±5 | 1.8±0.3 |
| 2007-01-16 17:58:43 | 30±8 | 2.2±0.5 |
| 2007-01-16 18:05:07 | 18±5 | 0.6±0.1 |
| **Average** | **24±2** | **1.8±0.1** |

Table 2: Field-aligned distance to the reconnection site during the February 2007 event as estimated from observed ion energy-pitch angle dispersions and a fit to Equation 1.

| Date | D [$R_S$] | T [hours] |
|---|---|---|
| 2007-02-01 18:02:18 | 60±60 | 7±6 |
| 2007-02-01 18:16:10 | 50±30 | 5±3 |
| 2007-02-01 18:23:06 | 30±20 | 3±1 |
| **Average** | **50±20** | **5±2** |
| | | |
| 2007-02-02 00:04:26 | 30±20 | 1.3±0.6 |
| 2007-02-02 01:08:26 | 33±9 | 2.2±0.5 |
| 2007-02-02 03:00:26 | 40±40 | 2±1 |
| **Average** | **37±9** | **1.7±0.3** |

# Figures

Figure 1: Cassini trajectory and mapped ionospheric footprint for the two case studies in this paper. Because the upstream conditions and hence the magnetospheric magnetic field are different in each case, two pairs of panels are shown, one for rev 37 (panels a and b) and one for rev 38 (panels c and d). Panel a) shows the trajectory of Cassini on rev 37 (red), with rev 38 (gray) for comparison, projected onto the X-Z plane in KSM coordinates. The field lines are traced using the Khurana et al. (2006) magnetospheric magnetic field model with the pressure set to match the estimated upstream conditions. Panel b) presents the mapped ionospheric footprint of Cassini on rev 37 (red) compared with rev 38 (gray) and a statistical UV auroral oval (Carbary et al., 2012). The footprint is mapped from Cassini's location by tracing the field lines in a simple field model consisting

of a ring current model (scaled to match the magnetopause subsolar distance) (Bunce et al., 2007) and third order internal field model (Cao et al., 2011). Panels c) and d) present the same information but for rev 38 (blue) with rev 37 (gray) for comparison and where the field models have been adjusted for the different upstream conditions. In all panels squares indicate the beginning of each day, and the bold segment indicate the intervals covered by figures 4 (panels a and b) and 11 (panels c and d).

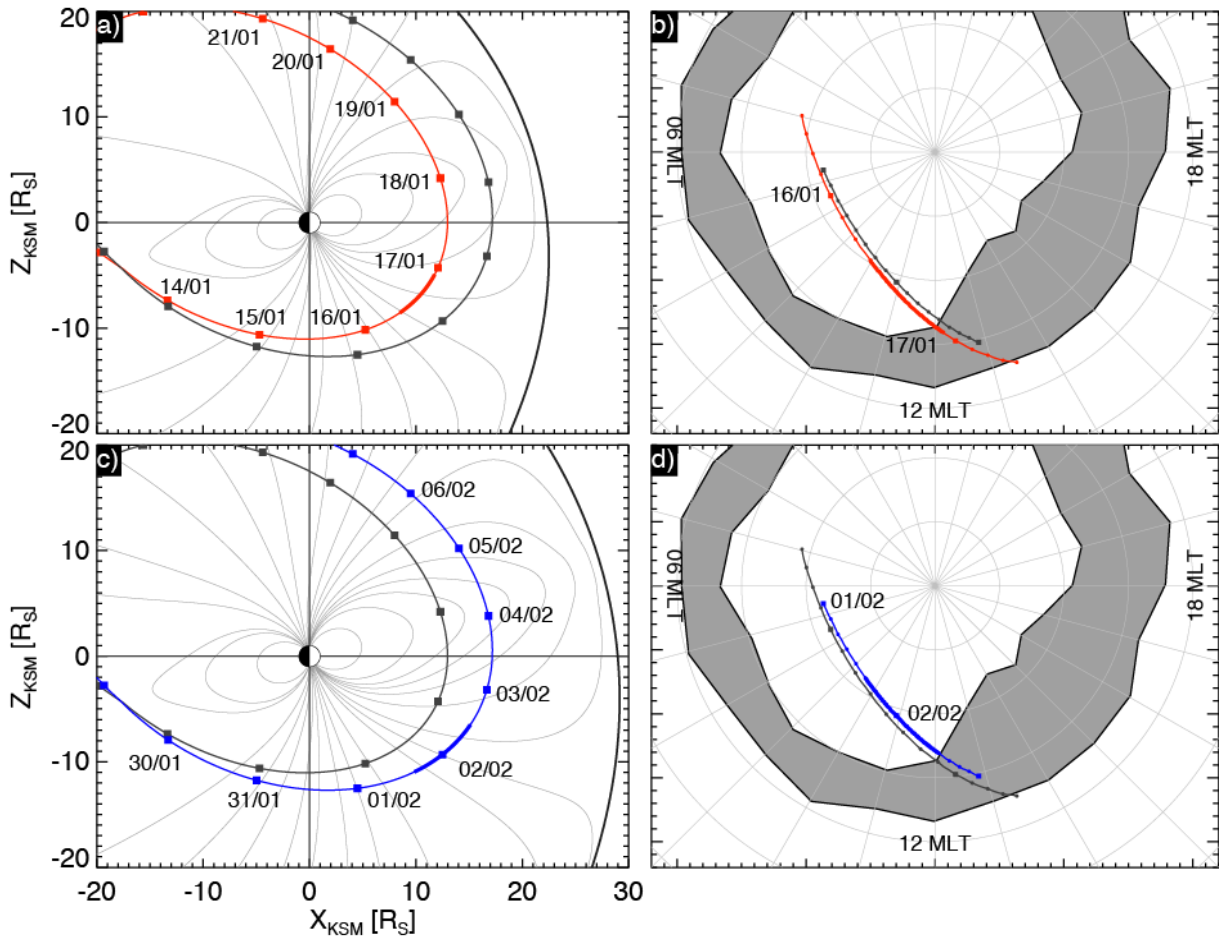

Figure 2: Overview of data for rev 37 (panels a-c) and rev 38 (panels d-f). Panel a shows energetic ion and electron differential number fluxes (DNF) [$cm^{-2}$ $s^{-1}$ $sr^{-1}$ $keV^{-1}$] measured by MIMI/LEMMS, panel b shows an omnidirectional electron spectrogram in units of differential energy flux (DEF) [$eV$ $cm^{-2}$ $s^{-1}$ $sr^{-1}$ $eV^{-1}$] and where the distributions have been filtered to remove low signal-to-noise bins and shifted to account for the spacecraft potential, panel c shows the field magnitude. Panels (d-f) show the same measurements for the rev 38 interval. Ephemeris data are shown below each set of panels and the bars/labels above each set of panels indicate the identified magnetospheric regions where PC indicates the polar cap, M indicates the magnetosphere, C/BL indicates the cusp/boundary layer, S indicates the sheath, and S/W indicates the sheath/solar wind.

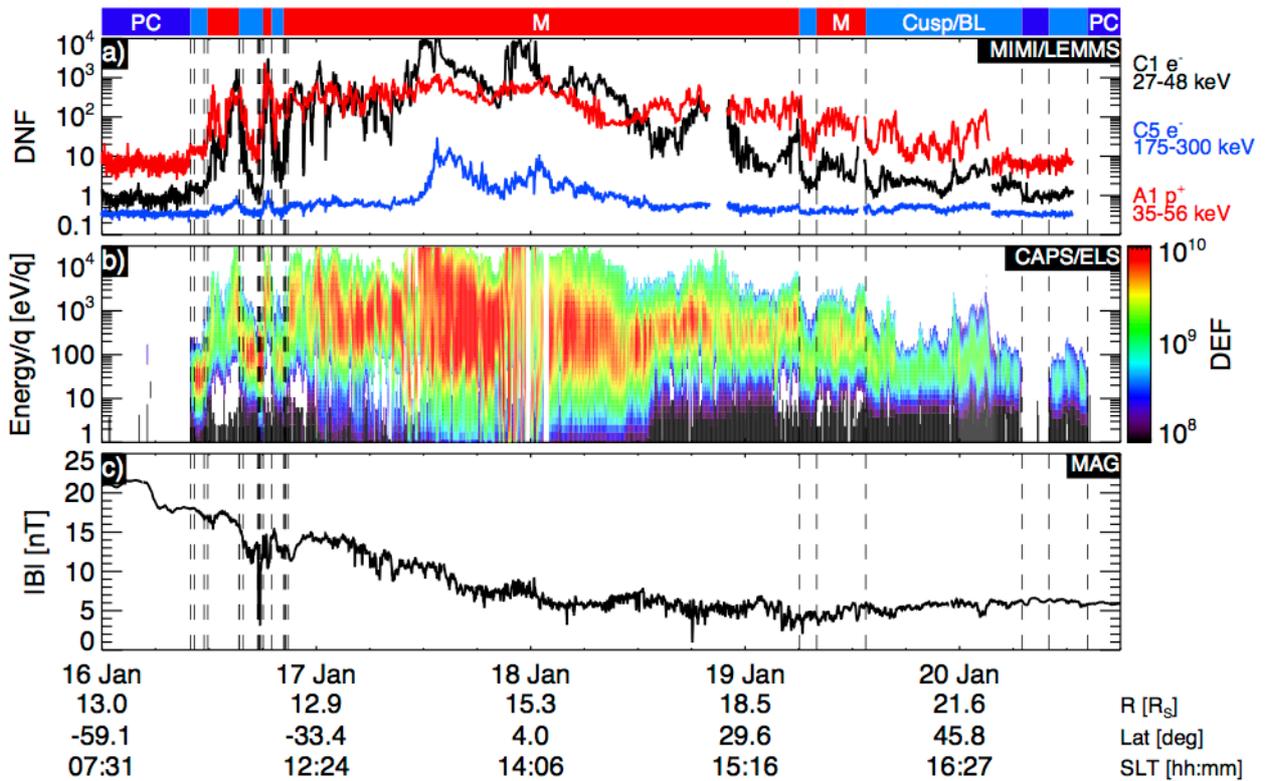

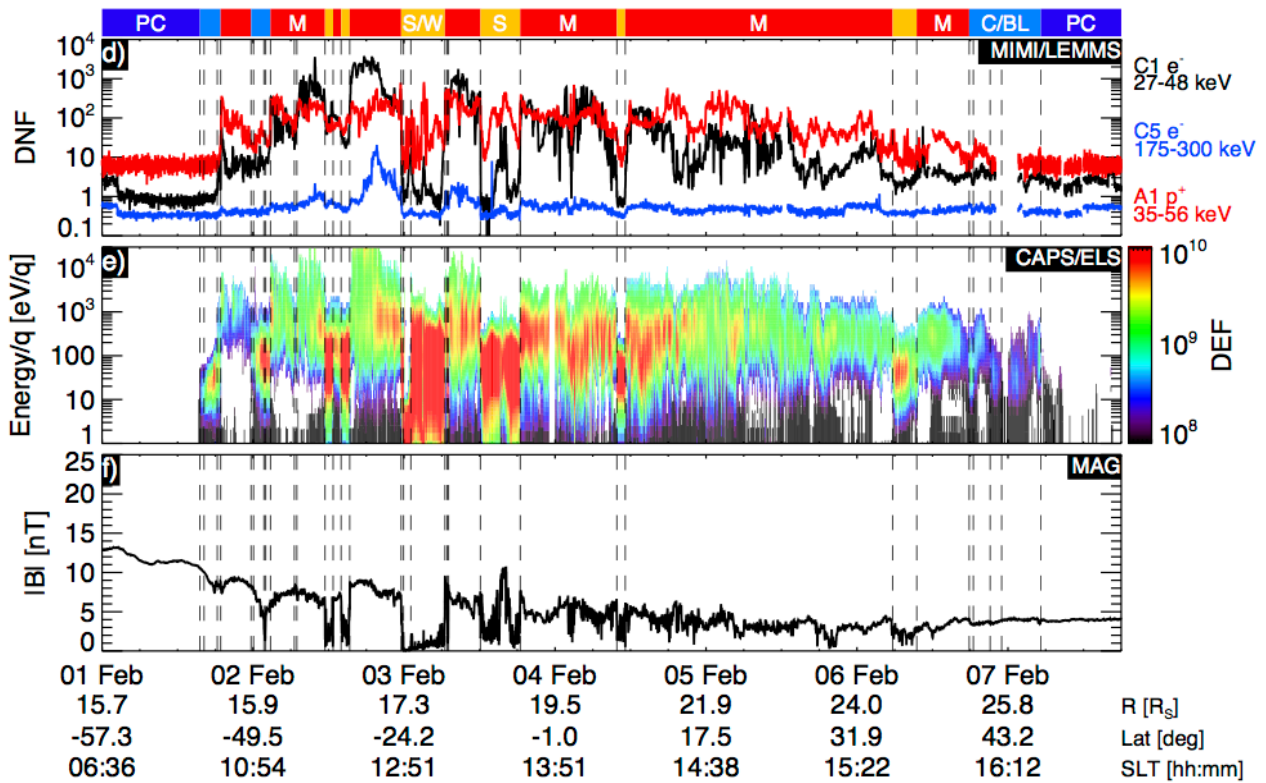

Figure 3: Upstream conditions during the January 2007 event. Panel (a) measured left-hand (therefore from the southern hemisphere) circularly polarized SKR power integrated between 3 – 1000 kHz (black) and 3 – 30 kHz (red). Panels (b) – (f) contain solar wind conditions propagated from 1 AU using the MSWiM model (Zieger and Hansen, 2008), showing b) the solar wind speed, c) the solar wind number density, d) the solar wind dynamic pressure, e) the tangential component of the IMF in the RTN (radial-tangential-normal) coordinate system, f) the IMF field strength. The tangential direction is approximately in the –y KSM direction, therefore negative $B_T$ is approximately duskward. The propagations have an arrival time uncertainty of ±15 hours therefore three time series are plotted, no lag (green solid curve), -15 hour lag (blue dash-dot curve), +15 hour lag (red dash-dot-dot-dot curve). The time series has also been lagged by +14 hours (black solid curve) to match the sharp increase in SKR power observed at 0600 UT on 16 January 2007 indicated by the vertical dotted line. The interval covered by figure 4 is indicated by the grey bars on each panel.

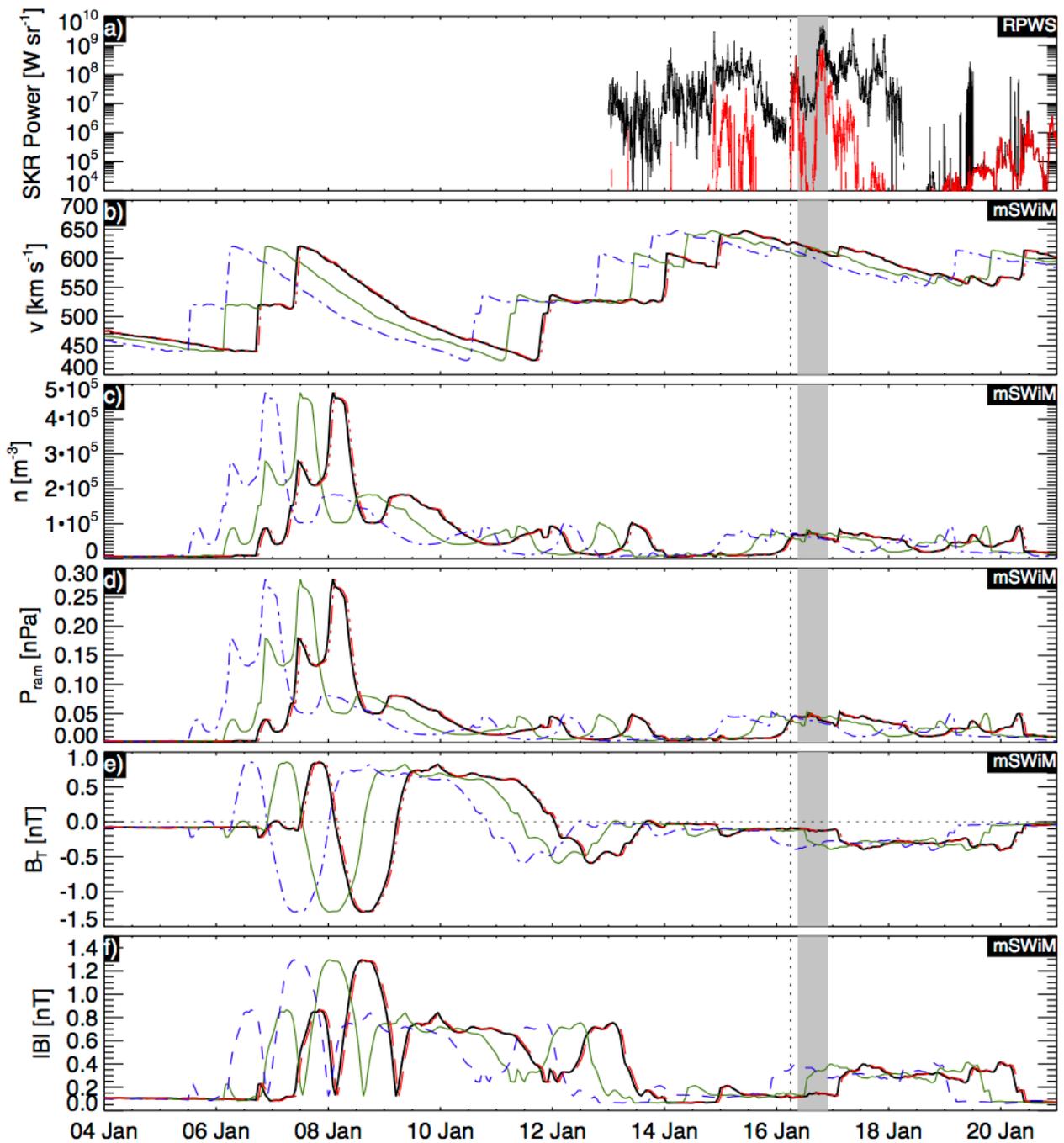

Figure 4: Overview of the January 2007 cusp crossing. The labels at the top of the figure show the identified regions: polar cap (PC), cusp, boundary layer (BL), and magnetosphere. Panel (a) shows a frequency-time spectrogram of the electric field from RPWS with the electron cyclotron frequency overlaid in white (calculated from the measured magnetic field strength); (b) an energy-time spectrogram of energetic electrons from LEMMS; (c) an energy-time spectrogram of plasma electrons from CAPS/ELS that has been filtered to remove bins with poor signal-to-noise and corrected for spacecraft potential; (d) an energy-time spectrogram of plasma ions from CAPS/IMS that has been filtered to remove bins with poor signal-to-noise; (e) relative abundances of plasma ions measured by the straight-through (ST) time-of-flight sensor in CAPS/IMS; (f) relative abundances of energetic ions measured by CHEMS; (g) magnetic field data in spherical polar coordinates (KRTP coordinates); (h) electron moments derived from CAPS/ELS where black lines are the number density, and blue is the temperature. Ephemeris information is provided at the bottom of the plot where invariant latitudes were estimated using a simple dipole plus current sheet model (Bunce et al., 2007; Cao et al., 2011).

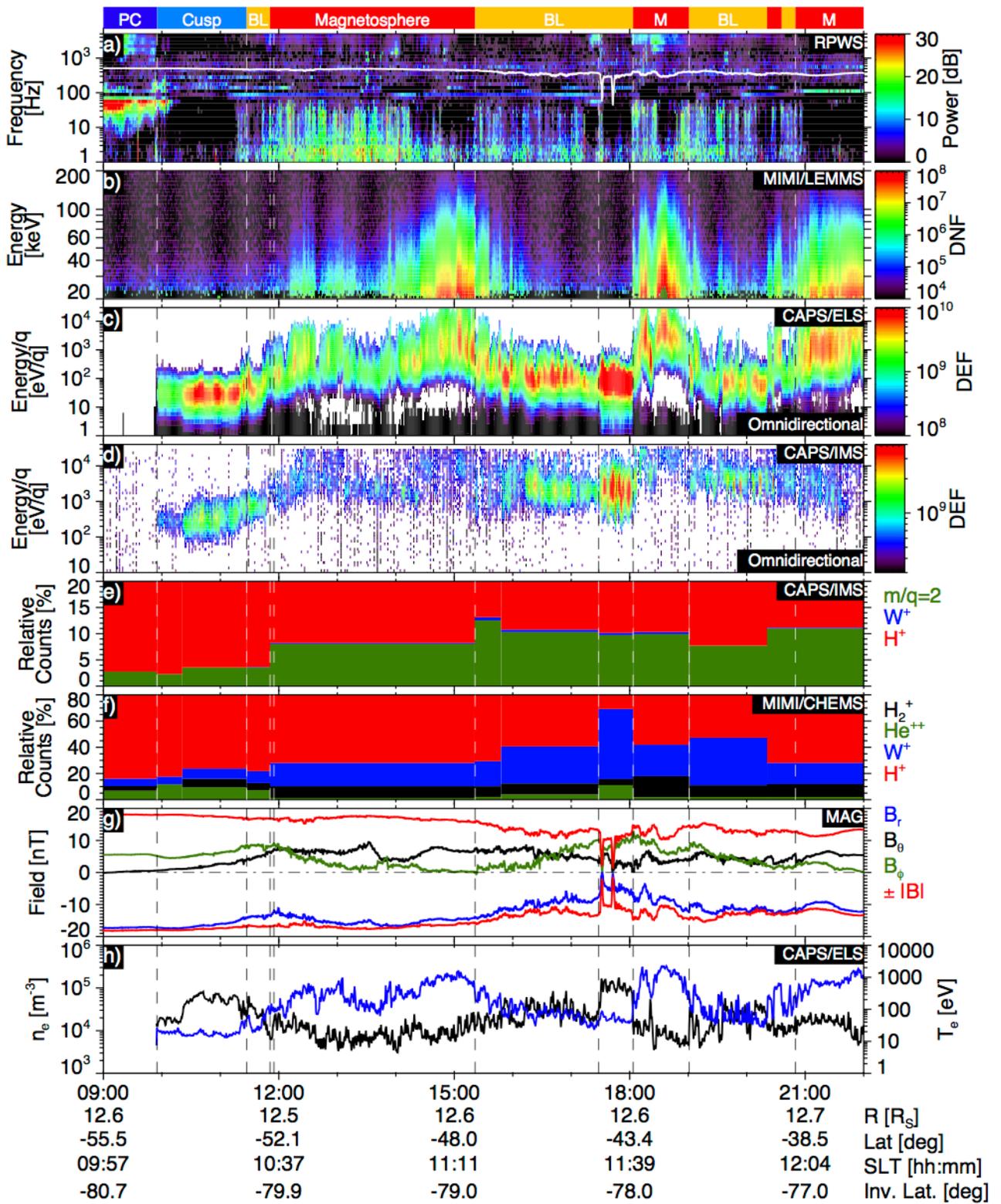

Figure 5: Ion counts measured by CAPS/IMS as a function of look direction around the spacecraft in a polar projection of the OAS coordinate system. The solid arcs show the boundaries between IMS anodes and the dashed lines show the center of the anodes. The orange circle shows the direction to the Sun, the green square shows the direction of ideal

corotation, red and blue triangles show 0º and 180º pitch angles. Ion counts are shown on a logarithmic scale from two IMS energy bins 724.1 eV/q (left) and 2.433 keV/q (right). Panels (a) and (b) show data from 17:34:42 to 17:38:25 UT, panels (c) and (d) from 17:51:46 to 17:54:57 UT, and (e) and (f) from 17:54:58 to 17:58:41 UT.

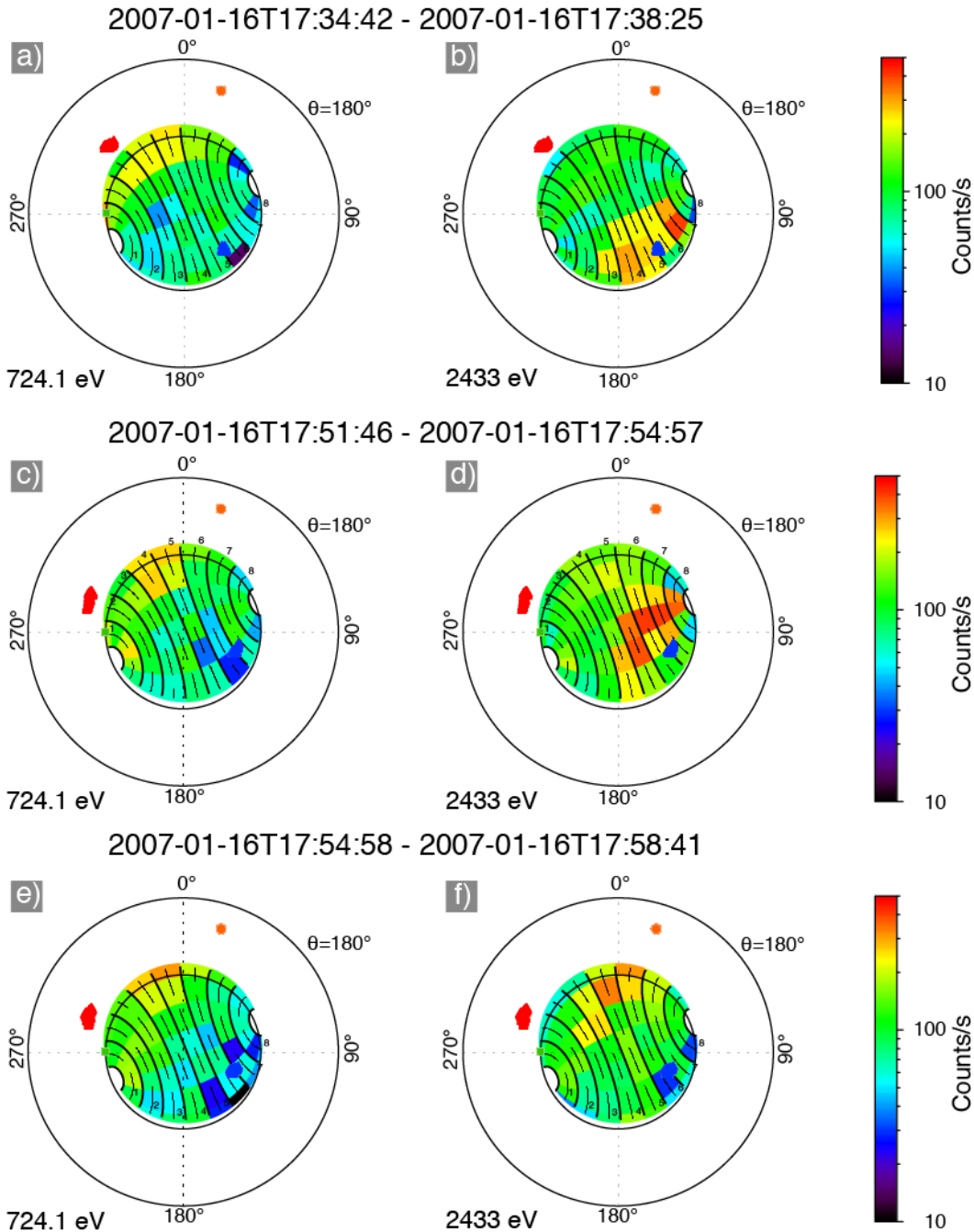

Figure 6: Measured energy-pitch angle ion dispersions during the 16 January 2007 event. In each panel we show a measured dispersion in differential energy flux (DEF), where the gray regions indicate no pitch angle coverage. The low energy ion cut-offs are automatically extracted by searching for when the ion flux drops below a signal-to-noise

ratio of 4. The uncertainty on this energy is taken as twice the energy resolution of IMS. The solid curve shows a fit of equation 1 to these ion cut-offs. Panels (a-d) show dispersions during the first interval in the cusp, and (e-h) show those during the boundary layer.

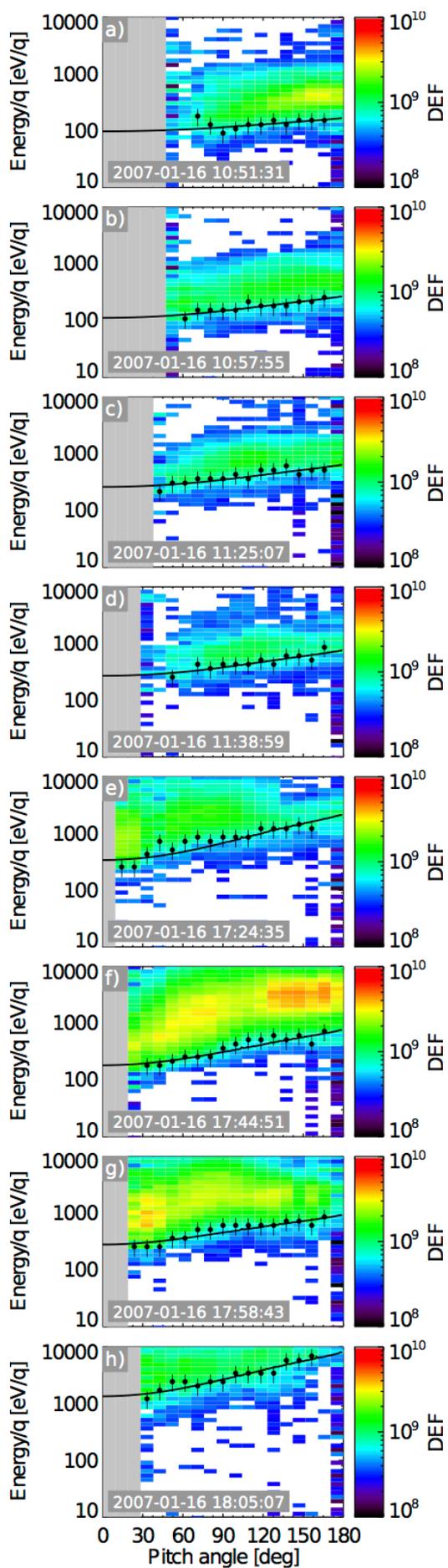

Figure 7: Cusp crossing data with Cassini's mapped invariant colatitude remapped into the rest frame of the auroral oval using the results of Nichols et al. (2008). Panel a) shows an energy-time electron spectrogram from CAPS/ELS, b) shows electron moments from CAPS/ELS where the black trace is the density and the blue trace the temperature, and c) shows the remapped invariant colatitude (black line) and the extent of the statistical auroral oval.

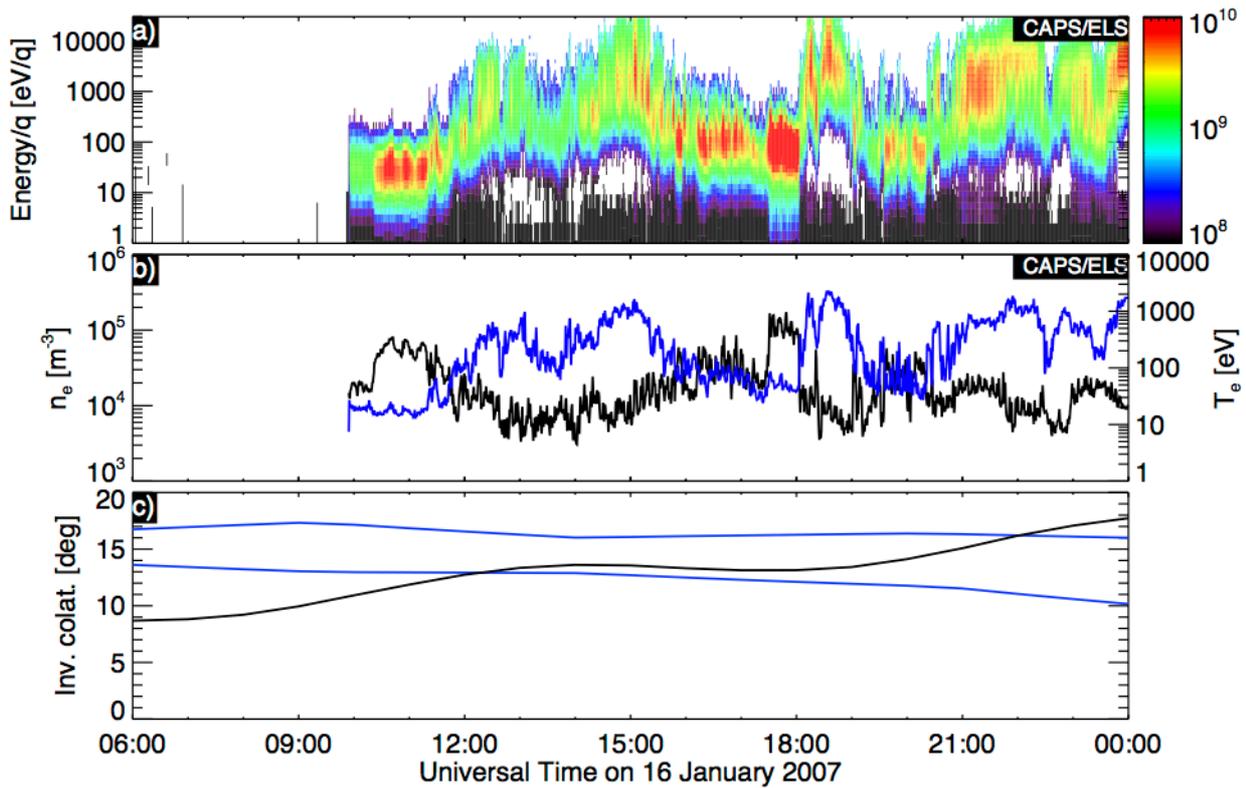

Figure 8: Upstream conditions during the February 2007 event in the same format as Figure 2. The solar wind time series has also been lagged by -16 hours (black solid curve) so that the increase in dynamic pressure corresponds to the magnetopause crossing observed at 1126 UT on 02 February 2007. The interval covered by Figure 9 is indicated by the grey bars in each panel.

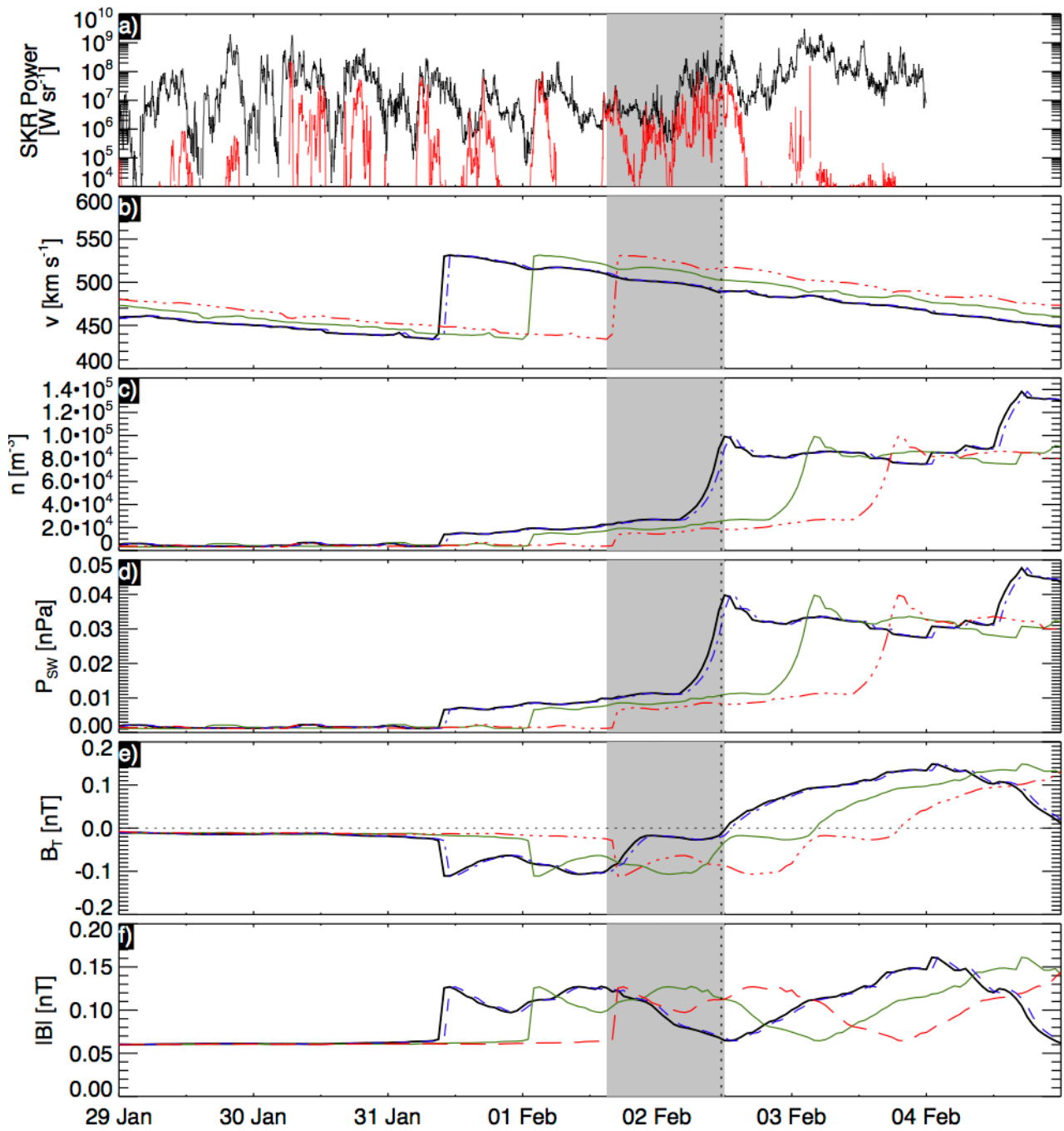

Figure 9: Overview of the February 2007 cusp crossing in the same format as Figure 4. The period in the magnetosheath is denoted by "MS" in the bars at the top of the figure.

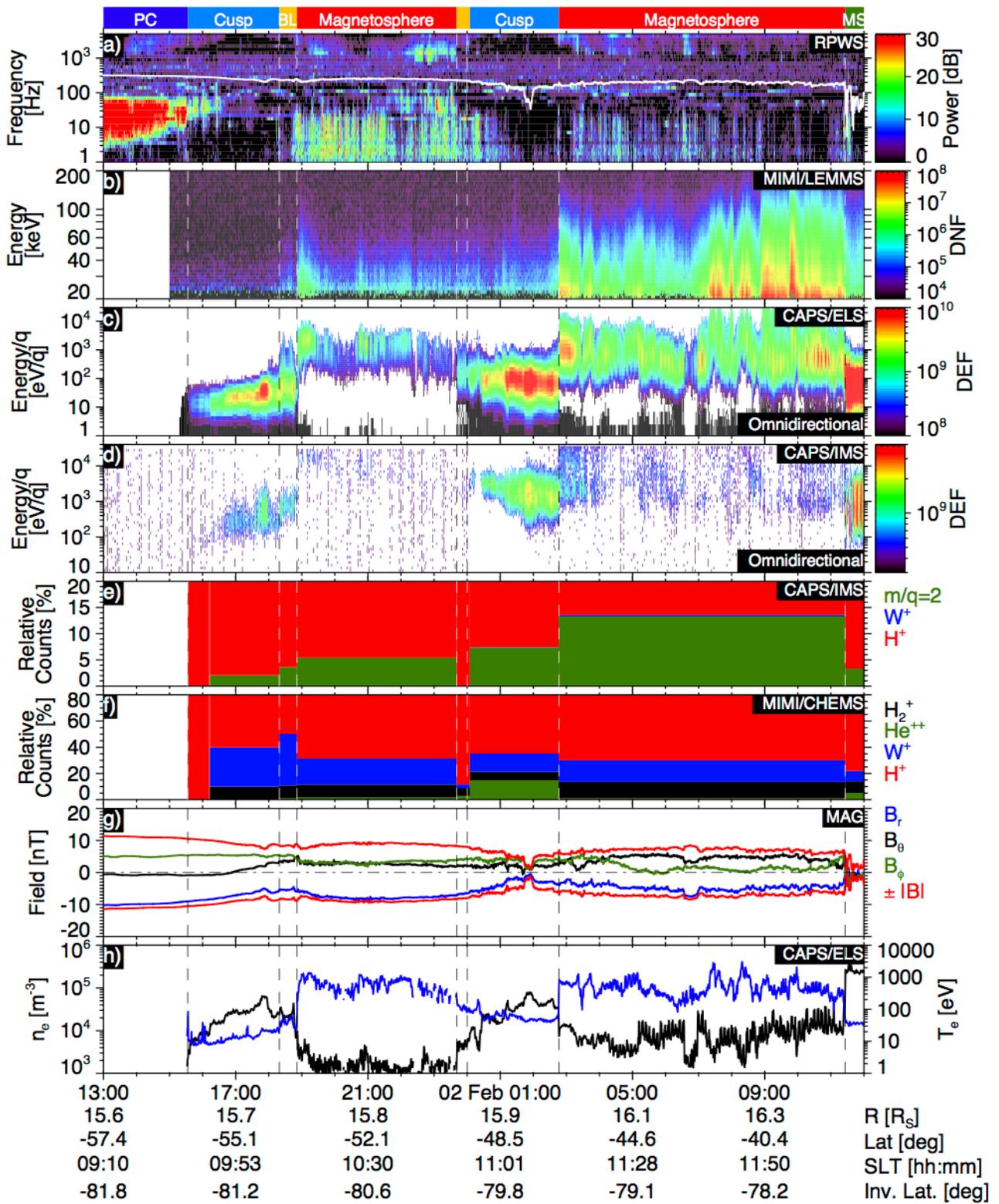

Figure 10: Measured energy-pitch angle ion dispersions during the February 2007 event in the same format as Figure 6. Panels a-c show dispersions during the first interval in the cusp, and d-e show those during the second interval in the cusp.

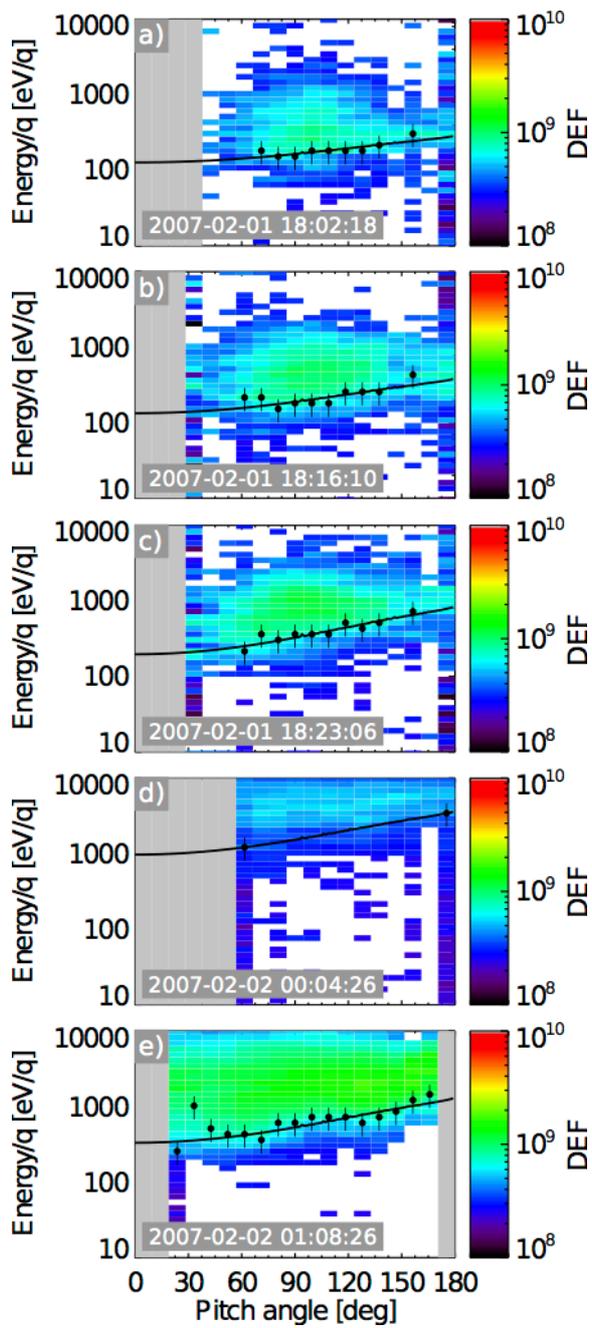

Figure 11: Cusp crossing data with Cassini's mapped invariant colatitude remapped into the rest frame of the auroral oval in the same format as Figure 7.

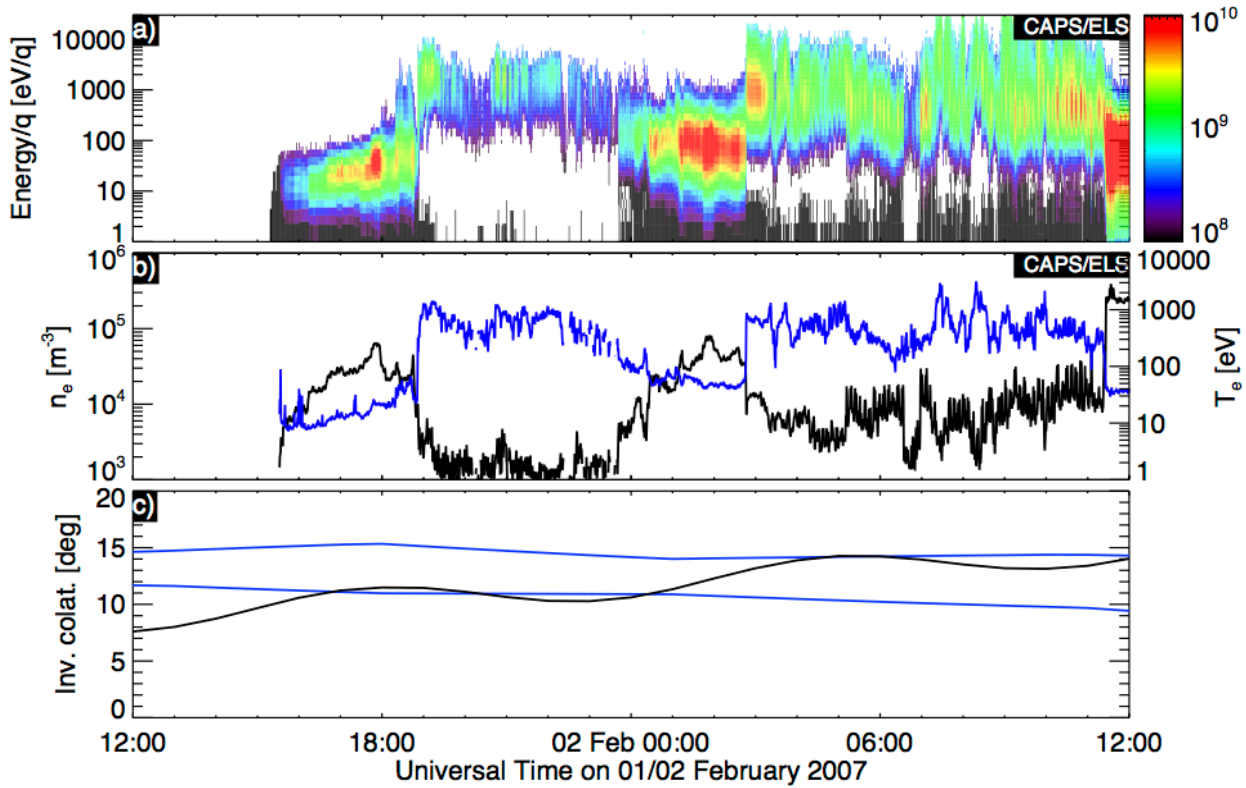

Figure 12: Projection of Cassini's trajectory onto the ionosphere (mapped using a simple field model (Bunce et al., 2007; Cao et al., 2011) which has been transformed into the rest frame of the auroral oval using the results of Nichols et al. (2008) and a magnetic field. This is done for both the 16 January 2007 (red) and February 2007 (blue) events considered in this paper. The statistical auroral oval from Carbary (2012) is included for reference.

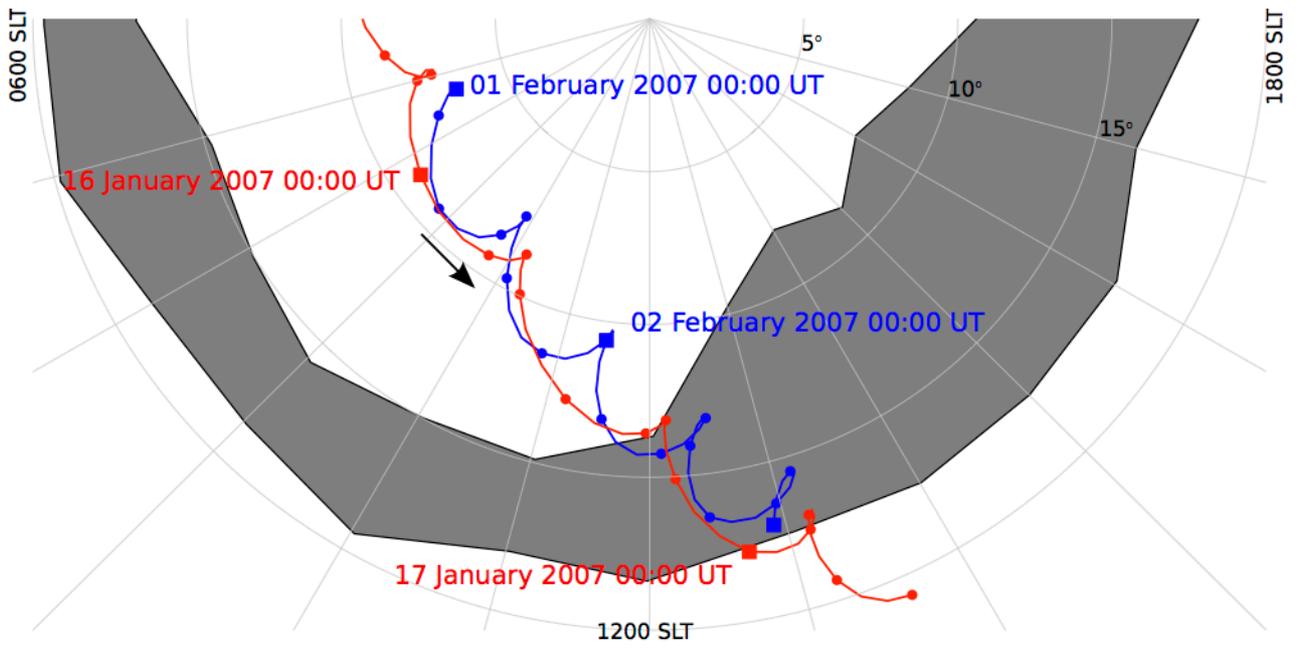